\documentclass[twocolumn]{aastex631}
\usepackage{float}
\usepackage{longtable}

\shortauthors{Noel et al.}

\begin{document}
\title{X-ray intraday variability of the TeV blazar Mrk 421 with  {\it XMM-Newton}}

\author[0000-0002-1568-6127]{A Priyana Noel}
\affiliation{Astronomical Observatory of the Jagiellonian University, Krak\'{o}w, Poland}
\email{anoel@oa.uj.edu.pl}

\author{Haritma Gaur}
\affiliation{Aryabhatta Research Institute of Observational Sciences (ARIES), Manora Peak, Nainital $-$ 263 001, India}

\author[0000-0002-9331-4388]{Alok C. Gupta}
\affiliation{Aryabhatta Research Institute of Observational Sciences (ARIES), Manora Peak, Nainital $-$ 263 001, India}

\author[0000-0003-4472-7204]{Alicja Wierzcholska}
\affiliation{Institute of Nuclear Physics Polish Academy of Sciences, PL-31342 Krakow, Poland}

\author[0000-0002-9199-7031]{Micha{\l} Ostrowski}
\affiliation{Astronomical Observatory of the Jagiellonian University, Krak\'{o}w, Poland}
  
\author[0000-0002-8105-4566]{Vinit Dhiman}
\affiliation{Aryabhatta Research Institute of Observational Sciences (ARIES), Manora Peak, Nainital $-$ 263 001, India}

\author[0000-0002-0705-6619]{Gopal Bhatta}
\affiliation{Institute of Nuclear Physics Polish Academy of Sciences, PL-31342 Krakow, Poland}

\begin{abstract}
\noindent
{Highly variable Markarian 421 is a bright high synchrotron energy peaked blazar showing wide featureless non-thermal spectrum making it a good candidate for our study of intraday flux and spectral variations over time.}
We analyse its X-ray observations {of over 17 years} taken with the EPIC-PN instrument to probe into the intraday variability properties. 
The photon energy band of 0.3 - 10.0 keV, and its sub-bands, soft 0.3-2.0 keV and hard 2.0-10.0 keV.
{To examine flux variability, fractional variability amplitude and the minimum variability timescale have been calculated. We also probed into the spectral variability by studying hardness ratio for each observation and the correlation between the two energy bands using 
discrete correlation function and 
inspecting the normalized light curves. The parameters obtained from these methods have been studied for any correlation or non-random trends.}
{From this 
work, we speculate on the constraints for possible particle acceleration and emission processes in the jet, for better understanding of the processes involving a turbulent behaviour except of shocks. A positive discrete correlation function between the two sub-bands indicates the role of the same electron population in the emission of photons in the two bands. The correlation 
between the parameters of flux variability and parameters of spectral variation and lags in sub-energy bands provide the constraints to be considered for any modelling of emission processes.}

\end{abstract}

\keywords{Galaxies: individual (\object{Mrk 421}) -- Galaxies: active -- Galaxies: jets -- Radiation mechanisms: non-thermal -- X-rays: galaxies}

\section{Introduction}
Blazars exhibit flux and spectral variability at all wavelengths of the observable electromagnetic (EM) spectrum, strong polarization ($>$ 3\%) from radio to optical bands and are accompanied with core dominated radio structures. Based on the unified model for radio-loud (RL) active galactic nuclei (AGNs), blazars emit the jets in the direction toward the Earth \citep{1995PASP..107..803U}. Blazars are a subclass of RL AGNs, which can be further split into the BL Lacertae (BL Lac) objects and the flat spectrum radio quasars (FSRQs). BL Lac objects show largely featureless composite optical to UV spectra and have predominant non-thermal  EM spectra while FSRQs show prominent emission lines. 

Multi-wavelength (MW) spectral energy distribution (SED) of blazars is characterized by broad double peaked structure in the logarithmic plot of $\nu f_{\nu}$ versus $\nu$ \citep{1997A&A...327...61G}. The low-energy SED peak situated from infrared to X-ray energies is due to the synchrotron radiation of relativistic non-thermal electrons in the jet. 
The high-energy peak ranges from MeV to TeV $\gamma-$ray energies and can be explained by leptonic or hadronic emission models. According to the leptonic model, the high energy emission is due to inverse Compton (IC) up-scattering of the synchrotron or the ambient photons by the highly relativistic electrons in the jet \citep[e.g.,][]{1998A&A...333..452K,2010ApJ...718..279G}. 
The hadronic emission models invoke either proton-photon cascade processes or synchrotron emission of extremely high energetic protons \citep[e.g.,][]{2003APh....18..593M}. 

Based on the location of the low energy peak in the SED, blazars are further classified into low-energy peaked blazars (LBLs) and high-energy peaked blazars (HBLs) \citep{1995ApJ...444..567P}. In LBLs the first peak of SED is visible in near-IR/optical frequency range, and in UV/X-rays photon energies in HBLs. The second peak lies in GeV $\gamma-$ray energies in LBLs and TeV $\gamma-$ray energies in HBLs.

Blazars display flux and spectral variability of the EM spectrum in a wide range of timescales, so to study them simultaneous MW observations are preferred .  Flux variations with time scales from a few minutes to less than a day are commonly known as a blazar micro-variability \citep{1989Natur.337..627M}, or an intra-day variability  (IDV) (as in \cite{1995ARA&A..33..163W}), or an intra-night variability (INV) (for example by \cite{2009MNRAS.399.1622G}). When the  variability timescale within several days to few months is studied it is referred to as a short term variability, while flux variations from months to years are described as a long term variability (like in \cite{2004A&A...422..505G}). 

Another  characteristic observed in blazars is the formation of loops in the Hardness Ratio (HR) - Intensity (I) diagrams, due to spectral hysteresis. As suggested \cite{1998A&A...333..452K}, there occurs a lag in photons between different energy bands because of difference in the acceleration and cooling time of electrons. \cite{2000aprs.conf..227S} pointed out  that the soft emission lag leads to clockwise loop while  hard photons lag results in an anti-clockwise loop in the HR-I  diagram.

In the present paper,  we study the detailed X-ray IDV properties of the blazar Markarian 421 using a long series of XMM-Newton satellite data. The blazar is also commonly known as Mkn 421 and  Mrk 421 ($\alpha_{2000.0} = \rm{11}^{h} \rm{04}^{m} \rm{27.2}^{s}$ and $\delta_{2000.0} = +\rm{38}^{\circ} \rm{12}^{'} \rm{32}^{"}$). Mrk 421 was first noted as a stellar-like object with a blue excess coinciding with a point-like bright nucleus of the elliptical galaxy. It was classified as a BL Lac object due to its featureless optical spectrum, compact radio emission, strongly polarized and variable fluxes in optical and radio bands \citep{1975ApJ...198..261U}. It is one of the nearest blazars at redshift $z =$ 0.0308 \citep{1975ApJ...198..261U} equivalent to the distance of 134 Mpc \footnote{Throughout this paper we use H$_{0} = \rm{71} \ \rm{km} \ \rm{s}^{-1} \ \rm{Mpc}^{-1}$, $\Omega_{m} = \rm{0.27}$, $\Omega_{\lambda} = \rm{0.73}$.}. By using spectra of its host galaxy, the central super massive black hole (SMBH) mass was evaluated to be (2$-$9) $\times$ 10$^{8}$ \rm{M}$_{\odot}$ \citep[e.g.,][]{2002ApJ...569L..35F,2002A&A...389..742W,2003ApJ...583..134B}. 
The synchrotron peak in the SED of Mrk 421 was found in X-ray energies higher than 0.1 keV, and thus the object was classified as HBL/HSP blazar. 
It was the first extra-galactic object from which TeV $\gamma-$ray emission was detected \citep{1992Natur.358..477P}. GeV  $\gamma-$ray emission was detected from Mrk 421 by {\it EGRET} (Energetic Gamma Ray Experiment Telescope) on board of {\it CGRO} (Compton Gamma Ray Observatory) satellite \citep{1992ApJ...401L..61L,1992IAUC.5470....2M}. By using observation of the source from various ground-based Cherenkov $\gamma-$ray telescopes, it is repeatedly observed as a TeV $\gamma-$ray emitting source \citep{1993AIPC..276..185S,2011ApJ...738...25A,2017ApJ...841..100A}. 

Mrk 421, as well as other TeV emitting blazars have been comprehensively observed in X-ray and $\gamma-$ray energies  to study its' flux and spectral variability properties \citep[e.g.,][and references therein]{1999ApJ...526L..81M,2000MNRAS.312..123M,2000ApJ...541..153F,2000ApJ...541..166F,2001A&A...365L.162B,2003A&A...402..929B,2005A&A...443..397B,2002ApJ...574..634S,2004A&A...424..841R,2007A&A...466..521T,2009A&A...501..879T,2010PASJ...62L..55I,2014A&A...570A..77P,2016ApJ...827...55K,2017ApJ...841..123P,2018MNRAS.480.4873A,2019ApJ...884..125Z}. This bright source is among the most extensively studied blazars in the observable EM spectrum due to its strong flux and spectral variabilities, and emission detected up to TeV energies. The object was a goal of several simultaneous MW observational campaigns over extended period of time \citep{2005ApJ...630..130B,2008ApJ...677..906F,2008A&A...486..721L,2009ApJ...691L..13D,2009ApJ...695..596H,2011ApJ...738...25A,2015A&A...576A.126A,2015A&A...578A..22A,2020ApJS..248...29A}.  In X-rays, Mrk 421 shows very complex flux and spectral variability \citep[see for review][]{2020Galax...8...64G}. MW observations of this source are used for fitting different emission models involving, e.g., the combination of SSC and IC processes, and occasionally lepto-hadronic models are applied \citep[see for review][]{2020Galax...8...64G}. 

It is known that many blazars are characterized with flux variations on IDV timescales in different EM bands  \citep[e.g.,][and references therein]{1989Natur.337..627M,2010ApJ...718..279G,2015MNRAS.451.1356K,2016MNRAS.462.1508G,2018Galax...6....1G,2018MNRAS.480.4873A,2019ApJ...884..125Z}. The most puzzling and not  yet well understood flux variations in blazars are those observed on the IDV timescales and most probably directly related to the activity in the close vicinity of the central SMBH of the blazar. In such a case, the flux variability can be applied to constrain the size and relativistic kinematics of emitting region, and the level of variability may strongly depend on the SMBH mass \citet{2004ApJ...617..939M}.  

To work on the intriguing issue of the blazar flux variability on IDV timescales, some of the us, involved in this collaborative work, initiated a pilot project in X-ray energies  based on the data of blazars taken from the public archives of various X-ray missions (e.g. XMM-Newton, NuStar, Chandra and Suzaku),  reported in a series of papers  \citep{2009A&A...506L..17L,2010ApJ...718..279G,2014MNRAS.444.3647B,2016NewA...44...21B,2015MNRAS.451.1356K,2016MNRAS.462.1508G,2017ApJ...841..123P,2018ApJ...859...49P,2018MNRAS.480.4873A,2019ApJ...884..125Z,2021ApJ...909..103Z,dhiman2021multiband}. 

Significant efforts were devoted by several researchers to study  XMM-Newton observations of Mrk 421 applying  various analysis techniques to the same observational data which we reanalyse in the present study.
\cite{2001A&A...365L.162B} published results on the first observation of Mrk 421 taken on 25 May, 2000, from XMM-Newton satellite. They reported that the source is more variable in higher energies than in lower energy bands, where the low energy band was taken below 1 keV and high energy band was taken above 3 keV. \cite{2018MNRAS.481.3563P} did a classification of AGNs on the basis of the X-ray variability in which they reported that for RL  AGNs, blazars,  dominates variability in higher energies over the lower energy range. 
\cite{2004A&A...424..841R} analysed the flaring in three observations of Mrk 421. They reported on the non-reliability of reporting a spectral lag using DCF (also in our work we note that DCF is not a robust way to study such lags in cases of light curves involving regular time trends). The HR-I diagram of the flaring part of the light curve for these three observations formed a loop, indicating spectral hysteresis. \cite{2018ApJ...864..164Y} analysed all the observations of Mrk 421 taken from XMM-Newton satellite. The authors did the fitting of flares of the light curves obtaining a power law distribution which is a signature of self organized criticality model that can explain flaring with involved magnetic reconnection processes.  
\\
Our new analysis of the {\it XMM-Newton}  extended series of Mrk 421 X-ray observations involves a multi-parameter study of the considered individual observational runs and, then, comparing all observations to reveal general trends (or lack of such trends) essential for understanding of the blazar X-ray emission generating processes. 
We utilise public archive data of 25 pointed observations of Mrk 421 with an EPIC-pn instrument on board of {\it XMM-Newton}  carried out within a period of $~$17 years (2000--2017) for analysis of flux and spectral variations on IDV timescales and to study the X-ray emission tentatively expected to be generated in the jet close to the central BH of the blazar. 
The paper is arranged as follows. In Section 2, we discuss the EPIC-pn public archive data  and its analysis. Section 3 provides brief description of the analysis techniques used in this study. In Section 4 the results of the data analysis are presented. A discussion and conclusions follow in sections 5 and 6, respectively.

\begin{table*}[htbp]
\label{table:observation table}
\caption{The analysed Mrk 421 observations from EPIC-PN instrument of {\it XMM-Newton}.}
\centering
\begin{tabular}{ccccccc}\hline \hline
Obs. ID    & Date of Obs. & Window mode & Obs. duration  & Pile up & Filter \\
           & dd$-$mm$-$yyyy   &            &  (ks)           &    &     \\\hline
0099280101 & 25$-$05$-$2000 & Small  & 66.4 & yes & Thick  \\
0099280201 & 01$-$11$-$2000 & Small  & 40.1 & yes & Thick  \\
0099280301 & 13$-$11$-$2000 & Small  & 49.8 & yes & Thick  \\
0136540101 & 08$-$05$-$2001 & Small  & 39.0 & yes & Thin 1 \\
0136541001 & 01$-$12$-$2002 & Timing & 71.1 & no  & Medium  \\
0158970101 & 01$-$06$-$2003 & Small  & 47.5 & yes & Thin 1  \\
0150498701 & 14$-$11$-$2003 & Timing & 49.3 & no  & Thin 1  \\
0162960101 & 10$-$12$-$2003 & Small  & 50.7 & yes & Medium  \\
0158971201 & 06$-$05$-$2004 & Timing & 66.1 & no  & Medium  \\
0153951201 & 07$-$11$-$2005 & Timing & 10.0 & no  & Thin 1 \\
0158971301 & 09$-$11$-$2005 & Timing & 60.0 & no  & Thick   \\
0302180101 & 29$-$04$-$2006 & Timing & 41.9 & no  & Thin 1  \\
0411080301 & 28$-$05$-$2006 & Small  & 69.2 & yes & Medium  \\
0411080701 & 05$-$12$-$2006 & Timing & 18.9 & no  & Medium  \\
0510610101 & 08$-$05$-$2007 & Timing & 27.6 & no  & Medium  \\
0510610201 & 08$-$05$-$2007 & Timing & 22.7 & no  & Medium \\
0502030101 & 07$-$05$-$2008 & Timing & 43.2 & no  & Thin 1 \\
0670920301 & 29$-$04$-$2014 & Timing & 16.2 & no  & Thin 1 \\
0670920401 & 01$-$05$-$2014 & Timing & 18.0 & no  & Thin 1  \\
0670920501 & 03$-$05$-$2014 & Timing & 18.0 & no  & Thin 1  \\
0658801301 & 05$-$06$-$2015 & Small  & 29.0 & yes & Thick  \\
0658801801 & 08$-$11$-$2015 & Small  & 33.6 & yes & Thick \\
0658802301 & 06$-$05$-$2016 & Small  & 29.4 & yes & Thick  \\
0791780101 & 03$-$11$-$2016 & Small  & 17.5 & no  & Thick  \\
0791780601 & 04$-$05$-$2017 & Small  & 12.5 & yes & Thick  \\\hline
\end{tabular}
\end{table*}

\section{XMM-Newton Observations and Data Reduction}

There are 55 pointed observations of Mrk 421 from May 25, 2000, to May 04, 2017, taken with EPIC-pn on board to {\it XMM-Newton} satellite. The best 25 observations selected for our analysis with a minimum duration of 10 ks are listed in Table \ref{table:observation table}.  Below, we provide information about {\it XMM-Newton} satellite, EPIC-pn data, and the performed data reduction and analysis.  
 
\subsection{XMM-Newton satellite}

XMM-Newton is an X-ray astronomy mission of European Space Agency launched in 1999.
The satellite has a 48 hour period of orbit revolution, with the perigee of 7000 km and the apogee of 114000 km.
It has 6 co-aligned instruments on board - three European Photon Imaging Cameras (EPIC), two Reflection Grating Spectrometers (RGS) and one Optical Mirror telescope (OM). 

EPIC is an X-ray instrument onboard XMM--Newton with one PN type CCD and the other two CCDs of MOS type, each with approximately 1500 cm$^{2}$ of geometric effective area and large FOV of 30$^{'}$ diameter. 
The photon energy range of the operation of EPIC instruments is 0.15$-$15 keV.
EPIC PN telescope has a higher time resolution than EPIC MOS telescope. 
The time resolution of 7 CCDs of PN telescope is 30 $\mu$s in timing and burst modes.  
The X-ray MOS CCDs are placed behind the telescope that contains the grating of the RGS. The grating redirects half of the incident flux to the RGS detectors, while 40\% of the incident flux is channeled towards the MOS cameras. {This makes PN instrument better for flux variability studies than MOS instrument.}

\subsection{Data selection}
The data comprising of the observations from PN camera were downloaded from the XMM-Newton Science Archive\footnote{http://nxsa.esac.esa.int/nxsa-web/$\#$search}. 
In our work,  to study IDV properties of Mrk 421 we require high quality observations, with each individual run longer than 10 ks ($\sim$ 3 hours) and only those have been selected for the analysis. 
 The data taken from PN instrument were selected because of its' higher sensitivity and lower photon pile-up effects as compared to the MOS instrument of EPIC. {This is the reason along with explanation given in section 2.1 that data only from PN instrument has been used. For the same reasons, other authors like \cite{2010ApJ...718..279G}, \cite{2014MNRAS.444.3647B}, \cite{10.1093/mnras/stab3738} and \cite{10.1093/mnras/stac286} have also used only PN data.}

{There were total 55 archived observations from PN instrument from the beginning of the satellite operation till June 2017. Out of all, 30 observations were excluded due to a number of quality and time reasons - spurious detections in 20 observations, no science products in 2 observations, no target visible in 5 observations and 7 had the observation time less than 10 ks. Mrk 421 has been observed to have a minimum variability time scale of 5.5ks from previous studies by \cite{Aggrawal_2018}, 1.1ks by \cite{https://doi.org/10.48550/arxiv.2102.00919} using \textit{Chandra} and \textit{Astrosat} data. Shorter observations than the variability timescale may not show variability properties thus observations greater that 10 ks were chosen to be able to study the variability property of the source in scales of minutes and few hours. Thus, after considering these conditions there were 25 observations left to accomplish our goal of doing intraday variability studies. These 25 observations as shown in \ref{table:observation table} were taken in small window mode or timing mode. The time resolution for small window mode for PN instrument is 5.7 ms while for timing mode is 0.03 ms. Thus, these observations are useful for studying variability over the scale of seconds.}

\subsection{Data Analysis}

The data analysis has been performed using the Science Analysis Software (SAS)\footnote{https://www.cosmos.esa.int/web/xmm-newton/sas-threads}.
The instrument science data, housekeeping and auxiliary files needed to perform the analysis are packed together as Observation Data Files (ODFs)
in the XMM-Newton data archive, where 
the Current Calibration Files (CCFs) are also available for data calibratation. 
To obtain the updated list of calibration files, a task \textit{cifbuild} was used and
the task \textit{odfingest}  to create a summary file of all the components in the ODF set.
The task \textit{epchain} consists of a set of tasks performed over PN data to produce an event list. 

Mrk 421 is a bright blazar generating photon pile-up effects. Pile-up effects are produced when two or more photons fall on the same pixel or neighbouring pixel and the detector counts it as one photon event with the sum of energies of falling photons. The model distribution of single photon events is compared to double photon event and if the two do not overlap then it signifies the presence of pile-up. If the both distributions are comparable then it means that there were no significant pile-up effects. To correct for the pile-up, the influenced by the effect circular core of the PSF is excised and an annular region without the pile-up is selected for analysis of this point source. The task \textit{epatplot} was used to check and correct it.\footnote{https://heasarc.gsfc.nasa.gov/docs/xmm/sas/USG/epicpileup.html} 

In the analysis we use the data from the limited energy range, from 0.3 keV to 10.0 keV, as there is known detector noise below 0.3 keV and  the measurements above 10.0 keV are subject to a high proton background caused by the solar activity \citep{Bulbul_2020, 2010ApJ...718..279G}. We monitored such effects by checking the derived light curves in this energy range. {For uniformity, all observations have been binned to 100 s, which helps to study variability in timescale of minutes.}

\section{Procedures applied in the analysis}

In this section we present the data analysis techniques and methods used in the paper to study individual observations' light curves (LCs). All the light curves were binned to 400 s for the data to be consistent throughout.
  
\subsection{Excess Variance}

One of the parameters we use to assess the Mrk 421 variability 
is the Excess Variance \citep{10.1046/j.1365-2966.2003.07042.x}.
It is derived from the difference between the total variance of the fluxes measured along the LCs and the total variance generated by the measurement errors. For $N$ being a number of the measured flux values, x$_{i}$, along the light curve and  having corresponding uncertainties $\sigma_{err,i}$  arising 
from the measurement errors the excess variance, $\sigma^2_{XS}$, is derived as

\begin{equation}
\sigma^{2}_{XS}=S^{2}-{\overline{\sigma^{2}_{err}}}
\end{equation}
where S$^2$ is the variance along the given observational sample LC,  given by
\begin{equation}
S^{2}={\frac{1}{N-1}\sum_{i=1}^{N}(x_i-\bar{x})^{2}},
\end{equation}
where $\bar{x}$ is the arithmetic mean of x$_{i}$, and the mean square error is
\begin{equation}
\overline{\sigma^{2}_{err}}=\frac{1}{N}\sum_{i=1}^{N}\sigma^{2}_{err,i}
\end{equation}
The fractional root mean square variability amplitude, denoted by F$_{var}$, is defined as a square root of the normalized excess variance 
\begin{equation}
F_{var}=\sqrt{\frac{S^{2}-{\overline{\sigma^{2}_{err}}}}{\overline{x}^2}}
\end{equation}
with an uncertainty of F$_{var}$ 
\begin{equation}
err(F_{var})=\sqrt{{\left(\sqrt{\frac{1}{2N}}\frac{\overline{\sigma^2_{err}}}{{\overline{x}^2}F_{var}}\right)}^2+{\left(\sqrt{\frac{\overline{\sigma^2_{err}}}{N}}\frac{1}{\overline{x}}\right)}^2}
\end{equation}

\noindent
This procedure has been used to quantify the flux variability of the source in the studied  total energy band (0.3 -- 10.0 keV), as well as in its'  hard (2.0 -- 10.0 keV) and soft (0.3-2.0 keV) sub-bands.

\subsection{Minimum timescale of the flux variability}

The minimum two-point flux variability timescale can be calculated as given by \citep{1974ApJ...193...43B} :

\begin{equation}
\tau_{var}= {\rm Minimum} \{ |\frac{{\Delta}t}{{\Delta}\ln{(x)}}| \}
\end{equation}

\noindent
where we select the minimum value from the series of all pairs (i, i+1) of the two successive data points, along the LC, $\Delta t_{i,i+1} = t_{i+1} - t_i$  and the respective $\Delta \ln{x} = (x_{i+1} - x_i)/x_i$. 
{
To account for effects of the uncertainties in the flux measurements in the calculation of variability timescales, in this formal derivations we  impose the condition that $|x_{i} - x_{i+1}| > \sigma_{x_{i}} + \sigma_{x_{i+1}}$ and treat all analyzed data uniformly irrespective of the possible gaps in the observations \citep[]{2018A&A...619A..93B,2019ApJ...884..125Z}).}
The uncertainties in $\tau_{var}$ are calculated using the respective count rates, $x_{j}$ and $x_{j+1}$, leading to the shortest variability timescale, with $\Delta x_{j}$ and $\Delta x_{j+1}$ being their corresponding uncertainties, as

\begin{equation}
\Delta\tau_{var} \simeq \sqrt{\frac{x_j^2\Delta{x_{j+1}^2}+x_{j+1}^2\Delta{x_j^2}}{x_{j}^{2}x_{j+1}^{2}{(ln[x_j/x_{j+1}])}^4}}\Delta t_{j,j+1}
\end{equation}

\noindent
The above weighted variability timescales were calculated using the light curves for the total energy range 0.3 keV to 10.0 keV.

\subsection{Hardness Ratio}

Hardness ratio (HR) is an efficient parameter to study the spectral variations of the source along its light curve, having an advantage over a full spectral modelling because of the possibility to study short time bins of the LC with low count statistics in each bin. 
We define HR in each analyzed time bin as a simple ratio of the photon counts in the selected hard (H) and the soft (S) energy bands \citep{2006ApJ...652..610P}.   
Physical meaning of the HR values depends on a number of  factors related to, which are usually poorly known, emission parameters of the source. 
One may expect the observed HR changes to describe the physical changes in the emitting source if the soft and hard ranges under study are related to different spectral components in the full spectrum. \\
\\
Following  other authors (e.g. \cite{2018MNRAS.481.3563P}), we decided  to select the energy scale of 2 keV to split the studied observational energy range into S (0.3 -- 2.0 keV)  and H (2.0 -- 10.0 keV) sub-ranges, to keep the hard energy range limited to relatively high energies, but also still to preserve reasonable number of high energy photons and limit HR statistical fluctuations.
To derive the HR in each time bin  along the LC we use photon counts, H and S,  in the respective energy sub-ranges:
\begin{equation}
HR=\frac{H}{S}
\end{equation}
where we use the \textit{lcurve} task of XRONOS program from HEASARC.

\subsection{Discrete Correlation Function}
When light curves are unevenly binned, the classical correlation function (CCF) technique, which is based on the interpolation in between the data points, is not precise in deriving correlations between compared light curves. Then a Discrete Correlation Function (DCF) technique \citep{1988ApJ...333..646E} can be used, similar to CCF, with additional advantage of not having to deal with binned data sampling and interpolating in time. DCF is employed here to check the correlation between the flux variabilities in H and S energy bands. In the applied procedure, at first, we compute the unbinned correlation, called unbinned discrete correlation function (UDCF), for the data in the aforementioned energy bands \citep{2017ApJ...841..123P} as

\begin{equation}
UDCF_{jk}=\frac{(S_j - \bar{S})(H_k - \bar{H})}{\sqrt{\sigma_S^2 \sigma_H^2}},
\end{equation}

\noindent
where S$_j$ and H$_k$ are "j-th" and "k-th" data points in the energy bands S and H, $\bar{S}$ and $\bar{H}$ are their mean values, and $\sigma_S^2$ and $\sigma_H^2$ are their variances, respectively. Each pair (S$_j$,H$_k$) is associated with a pairwise time lag $\Delta t_{jk} = t_{j} - t_k$.
The time binning of UDCF is done with a value depending on the bin time of the light curves in both the energy bands.  
To calculate DCF for each time lag $\tau$, we used the average result of UDCF.  Averaging is performed over 
all $UDCF_{jk }$ satisfying the condition $\tau - (\Delta\tau/2) \leq  \Delta t_{jk} \leq \tau + (\Delta\tau/2)$.

\begin{equation}
DCF(\tau)=\frac{1}{N}\sum  UDCF_{jk} .
\end{equation}

\noindent
where the summation is taken of the values of UDCF calculated over all permutations of j and k.
One should note that in general DCF between two light curves is not normalized to 1. 

To find the respective time lag, $\mu$, between the studied energy bands  we fit the Gaussian function to the DCF plot near its maxima \citep{2019ApJ...884..125Z} 
\begin{equation} 
DCF(\tau) = A e^{-(\tau- \mu)/2\sigma^2} 
\end{equation}
where $A$ is the maximum value of the DCF; $\mu$ is the time lag at which DCF peaks and $\sigma$ is the fitted width of the Gaussian function. For each DCF we select a limited range of $\tau$, where the maximum structure allows for reasonable Gaussian fit.

\subsection{Normalized H and S lightcurves}

The photon counts in H and S bands were normalized by dividing flux values (photon counts) in each time bin with the  average photon count in the respective band \citep{1997ApJ...486..799U}. The normalized light curves are plotted simultaneously against each other to understand trends of hard vs. soft X-ray photons in time, to independently evaluate the integral information provided by DCF, in particular the derived time lag of photons in the two energy bands. Visual comparison of the normalized light curves reveals real lags and non-uniformities between individual LC fluctuations. 

\subsection{The source variability duty cycle}

Estimation of the source variability duty cycle (DC) is commonly used for estimation of the time fraction in which the source exhibits time variability. We estimated the DC for Mrk 421 using the standard approach of \citet{1999A&AS..135..477R}, which was later used by many group of authors \citep[e.g.,][and references therein]{2016MNRAS.455..680A,2018MNRAS.480.4873A},

\begin{equation} 
DC = 100{\%} \cdot \frac{\sum_\mathbf{i=1}^\mathbf{n} N_i \Delta t_i}{\sum_\mathbf{i=1}^\mathbf{n} \Delta t_i} 
\end{equation}

\noindent
where we use the time duration, $\Delta t_i$, of the $i^{th}$ observation run (Observation ID ) and $N_i$ takes the value 0 for no IDV detection and 1 whenever IDV is detected, with the condition  F$_{var} > \rm{3} \times \rm({F}_{var})_{err}$ for detection of the IDV variability. 

\section{Results}

The selected 25 pointed X-ray observations of Mrk 421 from the EPIC-pn instrument onboard {\it XMM-Newton} provides us with an opportunity to carry analysis of flux variability, spectral variability, and cross correlated studies of soft and hard X-ray bands of the blazar on IDV timescales over ~17 years of observations. For individual Observational IDs the observation times in our sample varied from 10 ks to 71.1 ks. The detailed observation log is given in Table \ref{table:observation table} and the set of analysis results in Table \ref{table 2}. Here we only present the plots obtained for observation 0099280101 and the plots for all other observations are provided in the online Figure Set.
   
\subsection{Description of individual observations}

To start let us analyse all individual  Observation IDs studying set of plots including the light curve in total energy band, the light curves in the soft and the hard bands, HR-I diagram, a plot of HR vs time, the normalized light curves and a plot of the derived Discrete Correlation function. Some extra plots are added for cases where we compare our results with some
earlier work.

\subsubsection*{Obs. ID 0099280101}

\begin{figure*}[h!]
    \centering
    \begin{tabular}{ccc}
 
(a) \includegraphics[width=50mm]{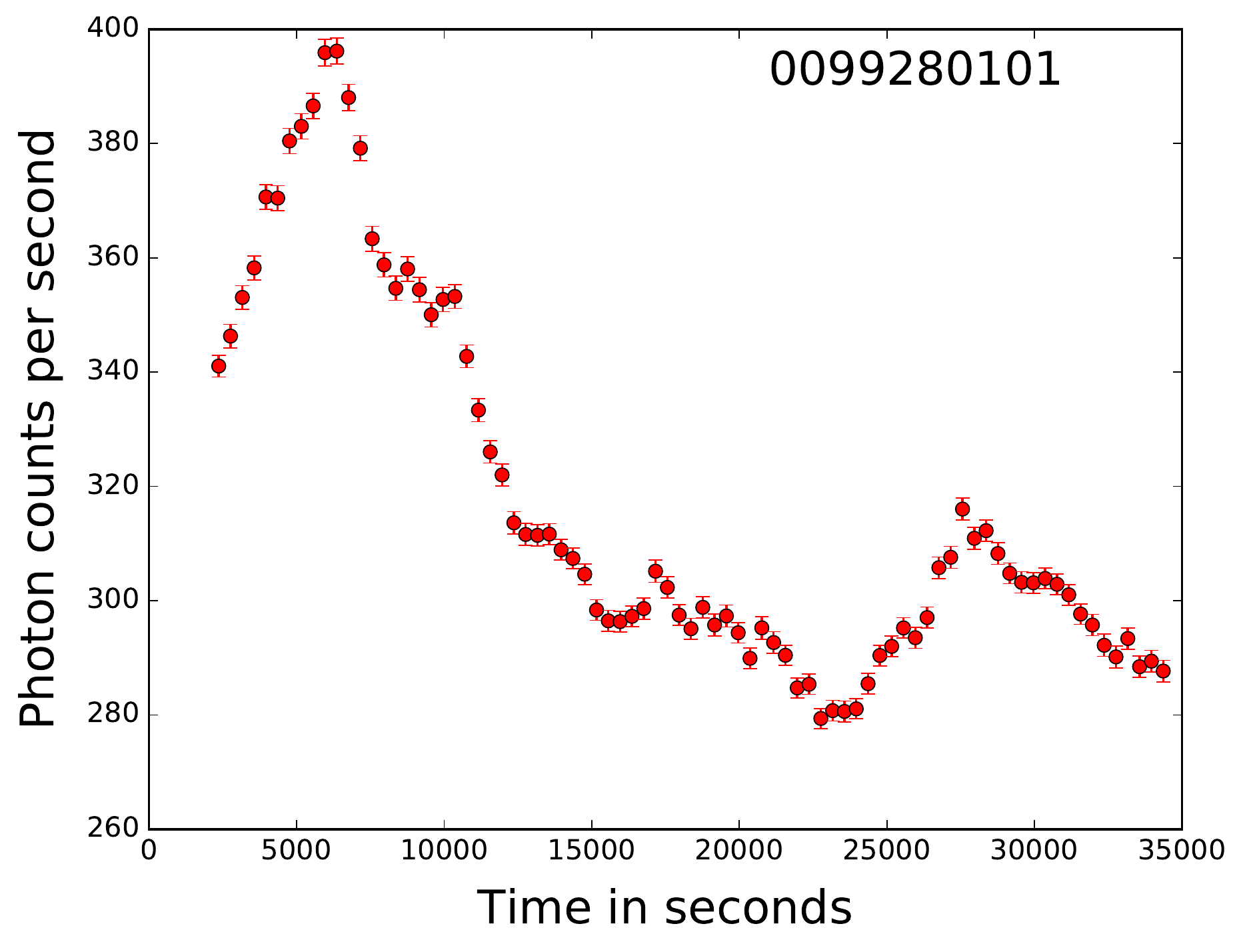} &
(b) \includegraphics[width=50mm]{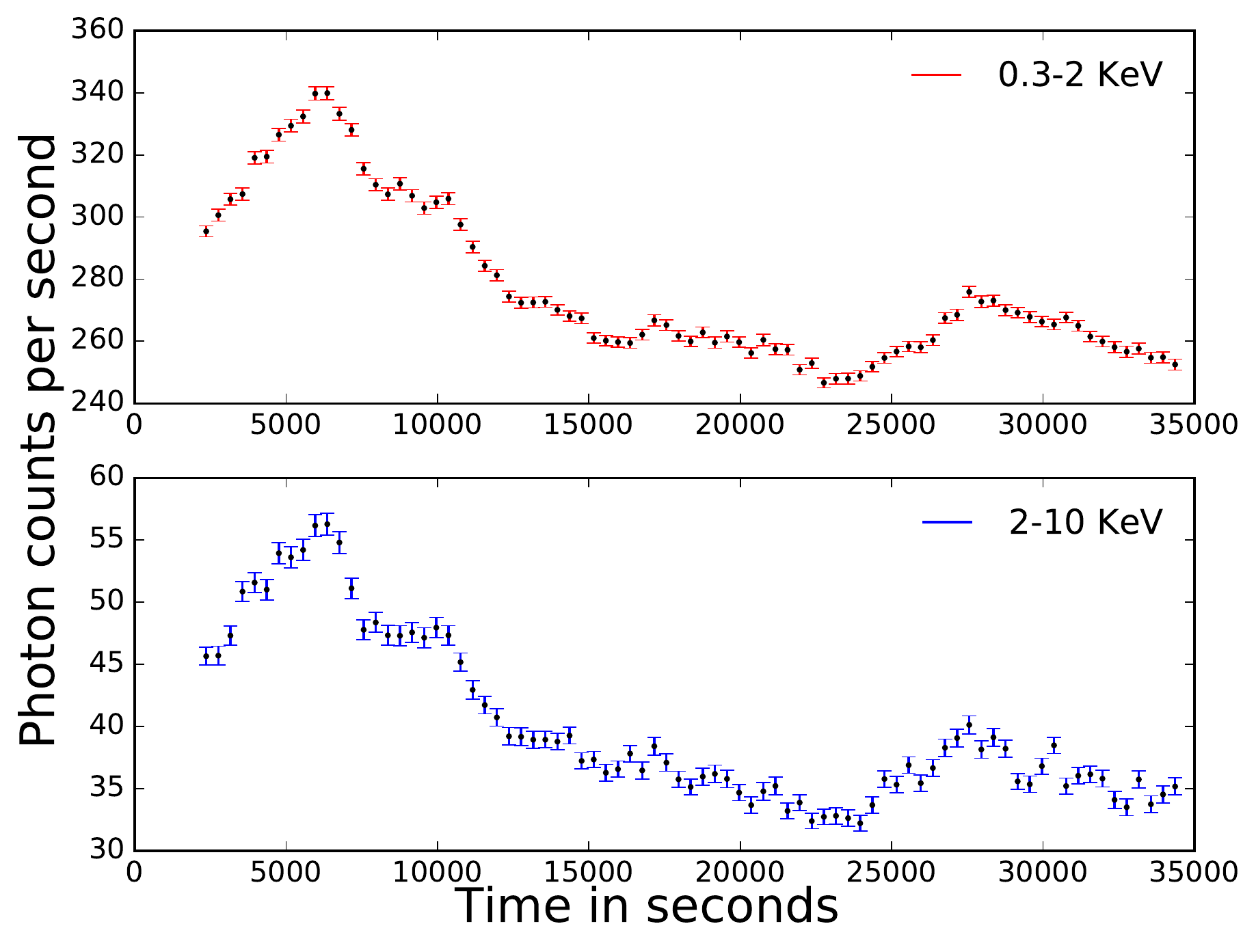} &
(c) \includegraphics[width=50mm]{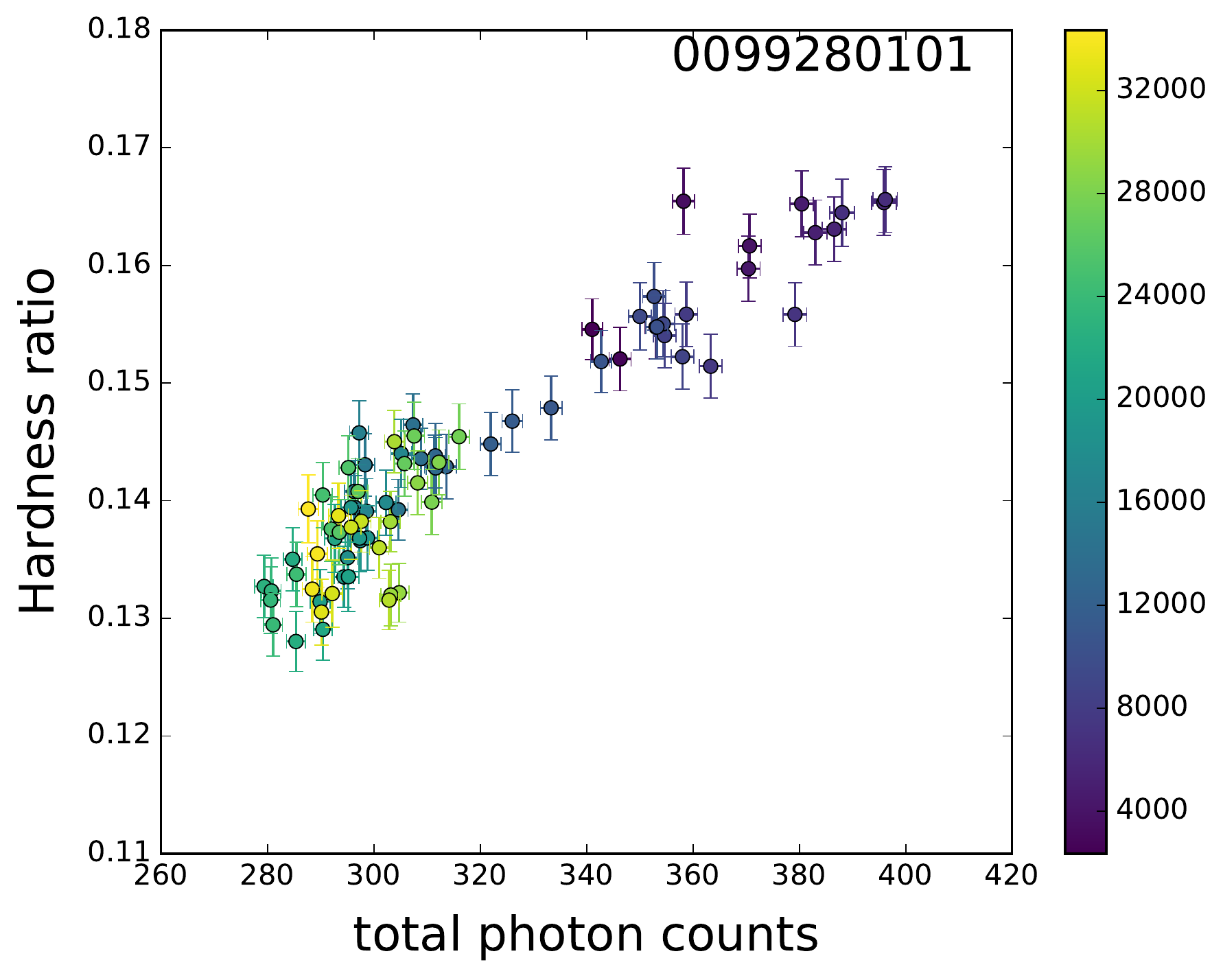} \\
 
(d) \includegraphics[width=50mm]{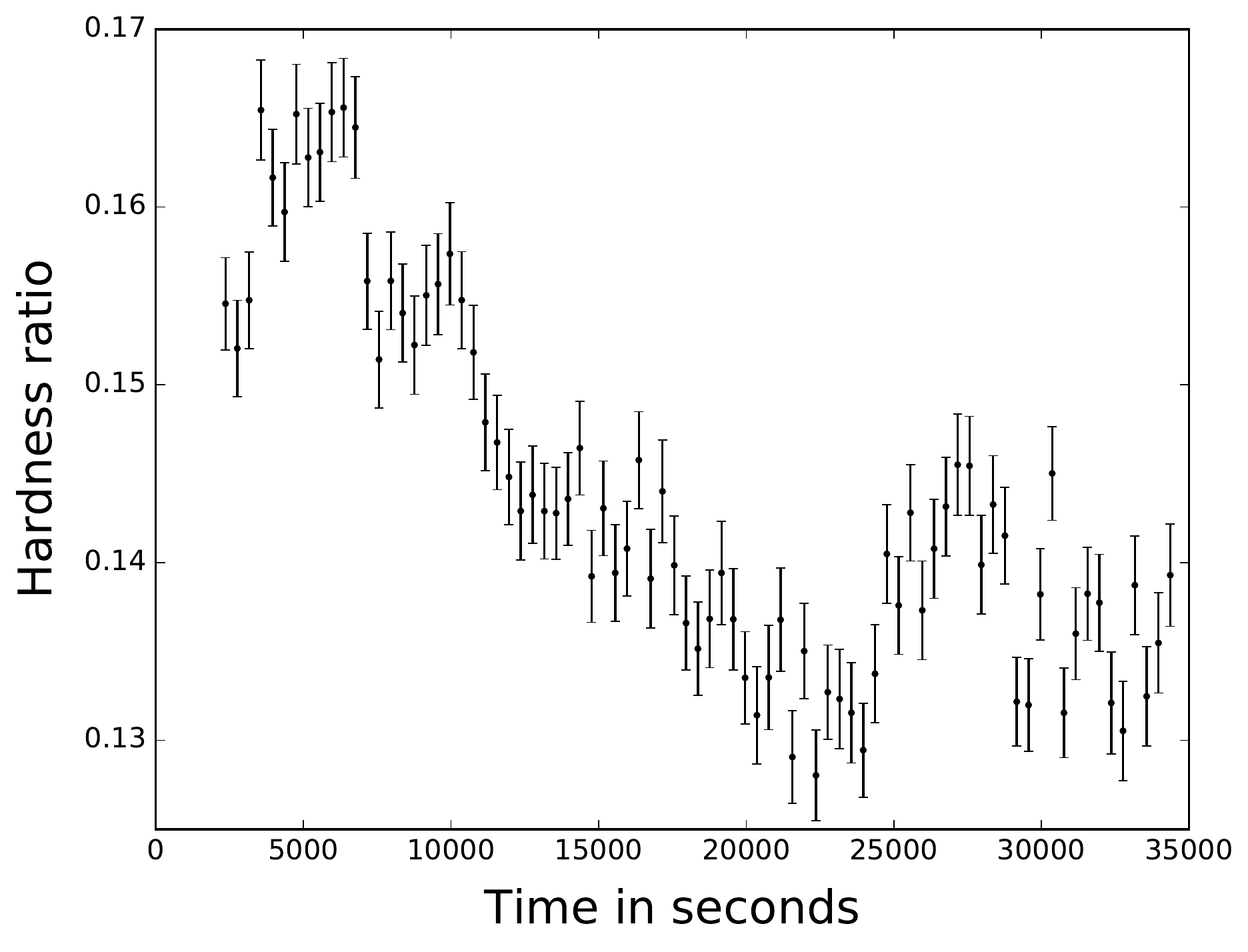} &
(e) \includegraphics[width=50mm]{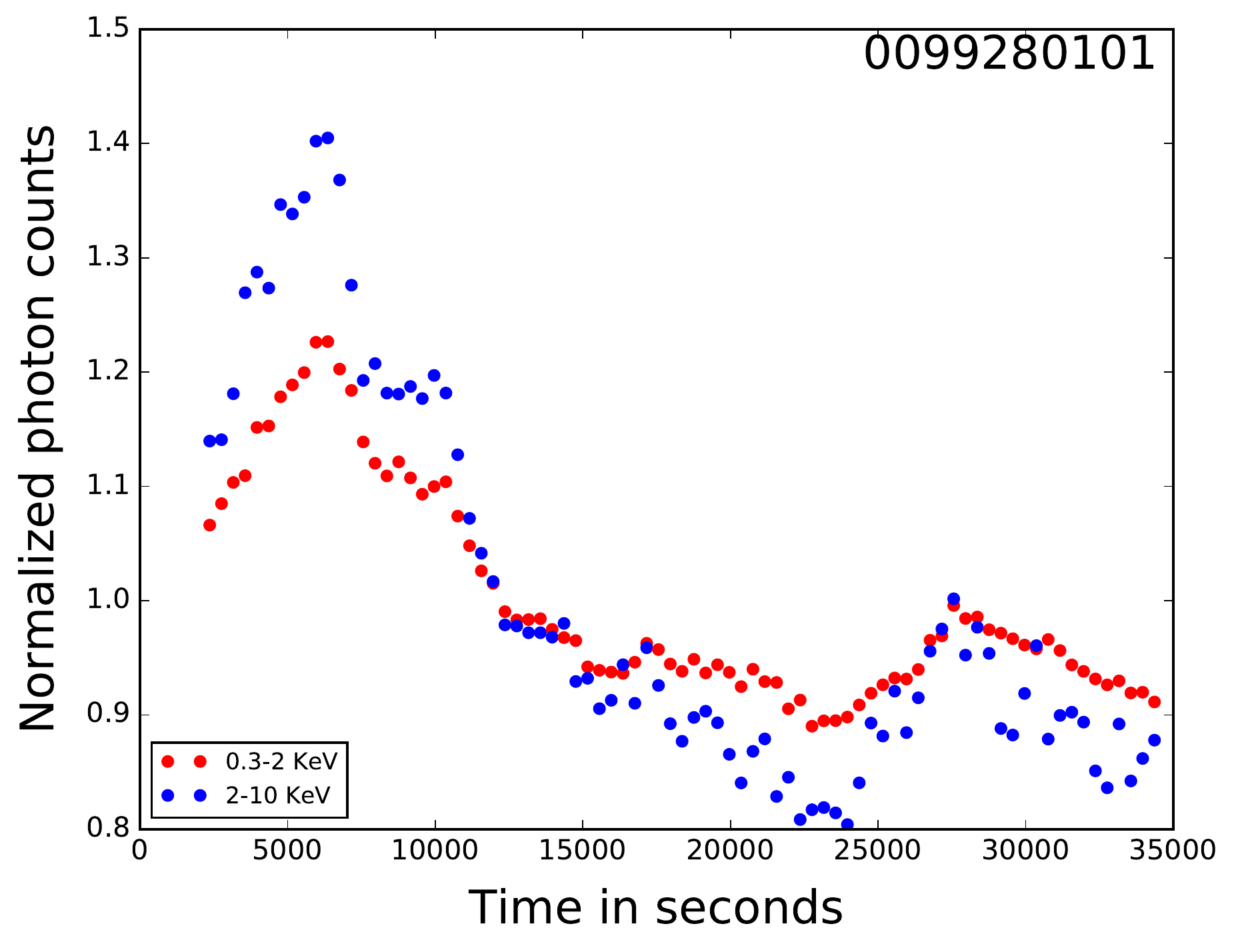} &
(f) \includegraphics[width=50mm]{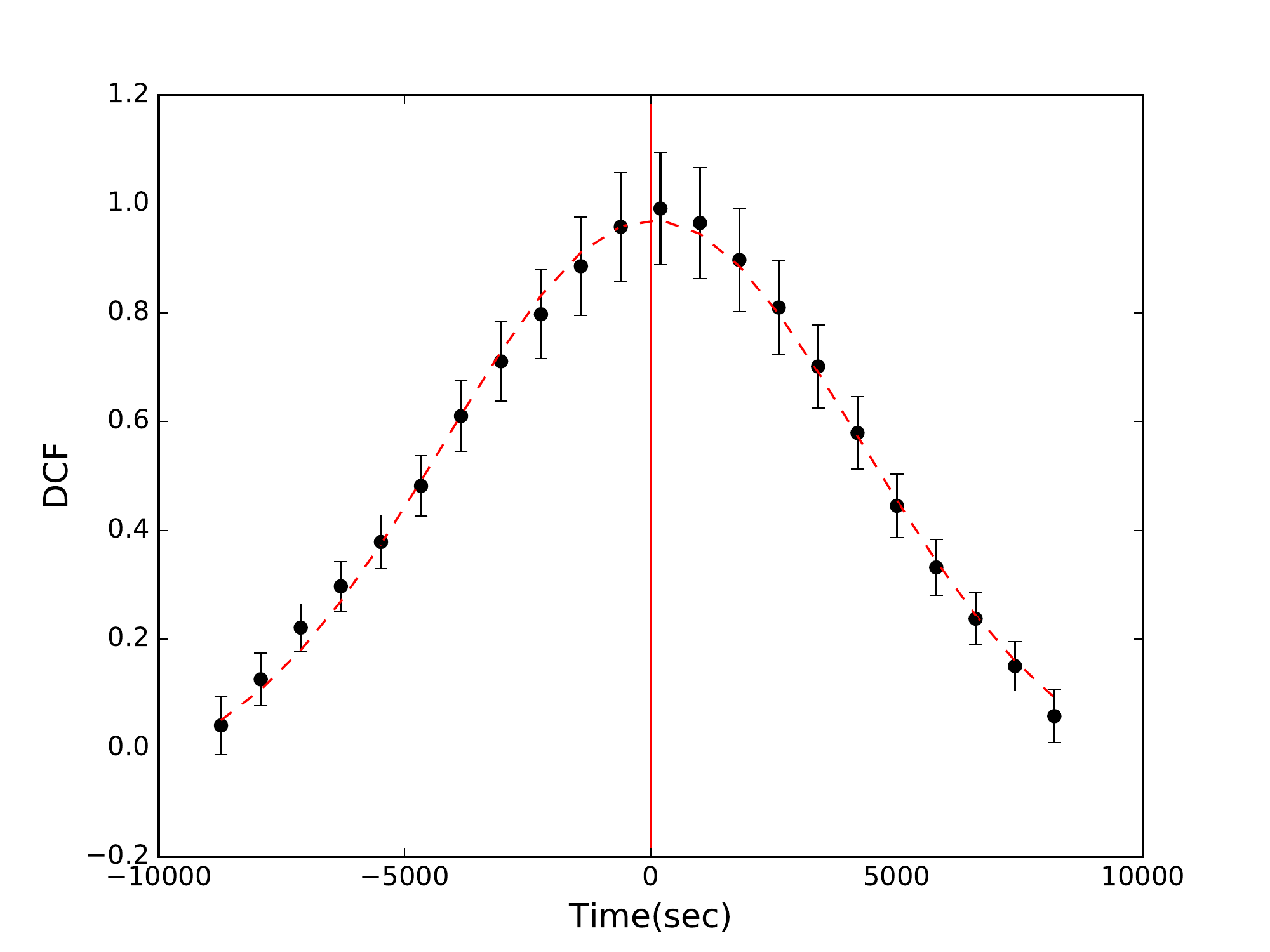} \\
 
\end{tabular}
\caption{ Obs. ID 0099280101 -- (a) Light curve in the total energy range (0.3-10.0 keV); (b) light curves in hard and soft energy bands; (c) hardness ratio vs. intensity (HR--I) diagram; (d) HR vs. time; (e) normalized light curves in the hard and the soft energy bands (points without error bars for the picture clarity); (f) a discrete correlation function (DCF). In panel (c)  the observational time {(in seconds)} is coded with a colour - dark blue for the beginning and yellow for the end of observation. 
The complete figure set (25 images) is available in the online journal.}
\label{fig 1}
\end{figure*}

The 
66.4~ks observation  of Mrk 421 
(see figure \ref{fig 1}) {display} 
high values of fractional variability of 10.16\%, 16.49\% and 9.25\% 
in {the} total, hard and soft energy bands, {respectively}. There are two visible flares in this observation, {one} 
from the beginning of the observation, 
and the other starting {at}
$\sim$ 24~ks. 
The variable HR of the emission shows a clear harder-when-brighter behaviour in the HR--I plot.  
There is no 
systematic trend 
seen in normalized light curves in {the} hard and {the} soft energy bands, where rise and fall of fluxes in both the bands 
look {similar}. 
Higher variability is visible in the hard band which is similar to the result obtained by \cite{2001A&A...365L.162B} . {Soft and hard band fluxes are well congruent with each other, thus inferring zero lag.}

\subsubsection*{Obs. ID 0099280201}









{The} 40.1~ks observation 
{displays} high variability of above 9\% in all considered energy bands (Table \ref{table 2}). 
There 
{are three distinct flares} visible 
{during} the observation.
{The hard} flux proceedings {soft} and spectral variability which can possibly contribute to the non-zero {lag} $\mu=2.44 $ ks observed in {the} DCF {analysis}. 
This is also clearly visible in the the normalized light curves where we see that after 20 ks of observation, there is a {steeper} rise in hard photons compared to the 
soft photons. There is a slight loop {structure} in {the} clockwise direction visible in HR-I diagram in the beginning of observation, while the overall HR distribution looks pretty flat, without any 
harder-when-brighter trends.

\subsubsection*{Obs. ID 0099280301}








The photon fluxes measured for this 49.8~ks observation ranges from 400 to 463, 51 to 84 and 451 to 548 respectively for soft, hard and total energy bands
{$F_{var}$ in hard energy band is more than three times higher than in soft and total energy band.}
In our view in the HR-I  diagram (not quite clear) indications of  two loops can be traced, one in anti-clockwise direction followed by the one in clockwise direction after around 20ks of observation. The value of fitted  DCF time lag of -0.19 ks is small. After inspection of panel (e) of fig. set 3, it is compatible with a zero lag.

\subsubsection*{Obs. ID 0136540101}







{Just like the previous observations, the variability in hard energy band is comparatively very high while for soft and total energy band, it is almost equal. The flux ranges from 329 to 423, 28 to 50 and 358 to 471 cts/s in soft, hard and total energy bands.}
The derived minimum flux variability timescale is ~3.4 ks. This behaviour of variability in short time intervals is also clearly visible in the normalized light curves in the panel (e). A non-significant time lag of -0.21 ks derived from DCF fitting indicates that the soft photons should precede the hard ones, but there is no such trend observed by eye in panel (e). The hardness-intensity plot depicts a clear harder when brighter behaviour, with imbedded loop like structures in the general trend. 

\subsubsection*{Obs. ID 0136541001}







\begin{figure*}[h!]
    \centering
    \begin{tabular}{ccc}
(a)\includegraphics[width=48mm, angle=0]{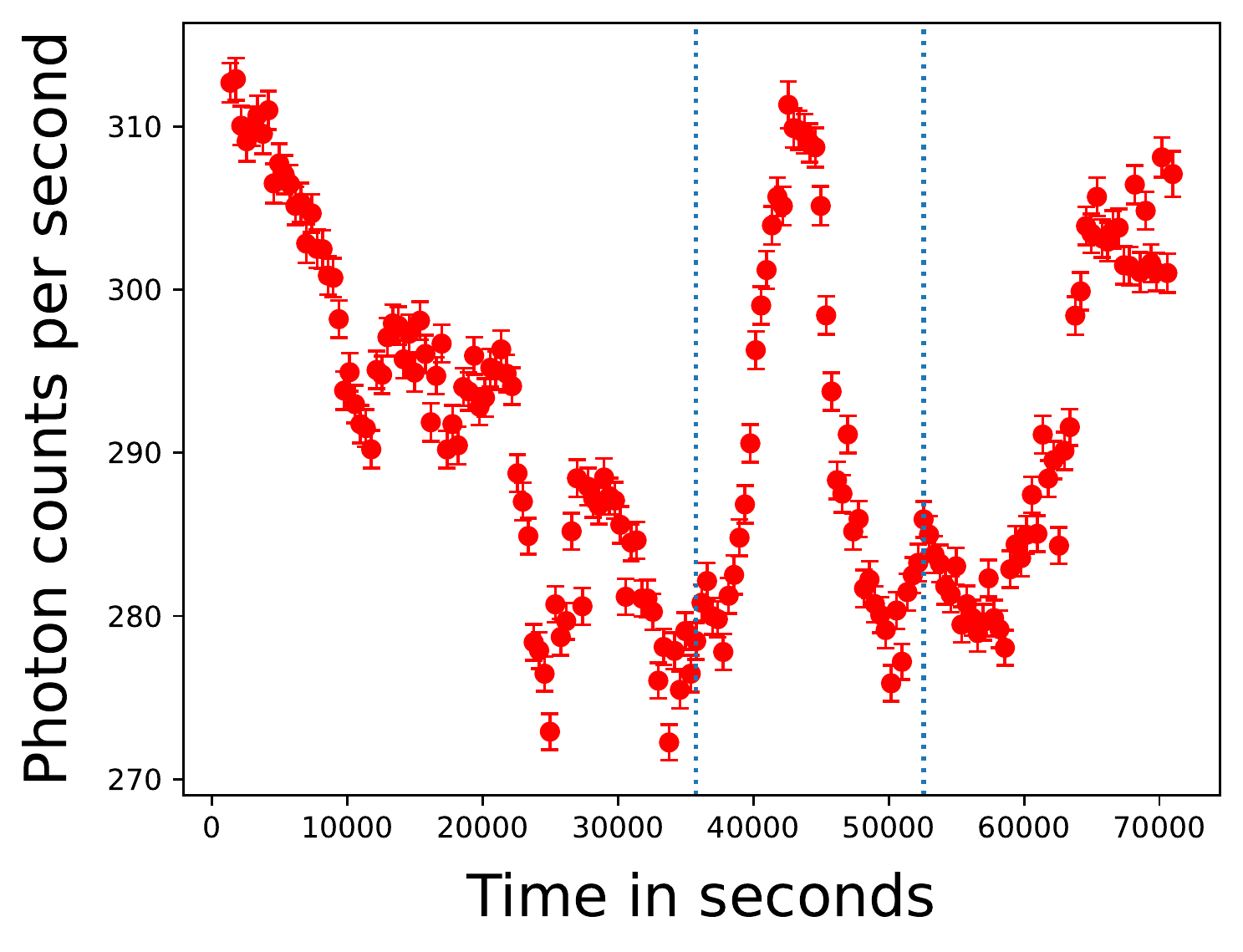} &
(b) \includegraphics[width=48mm, angle=0]{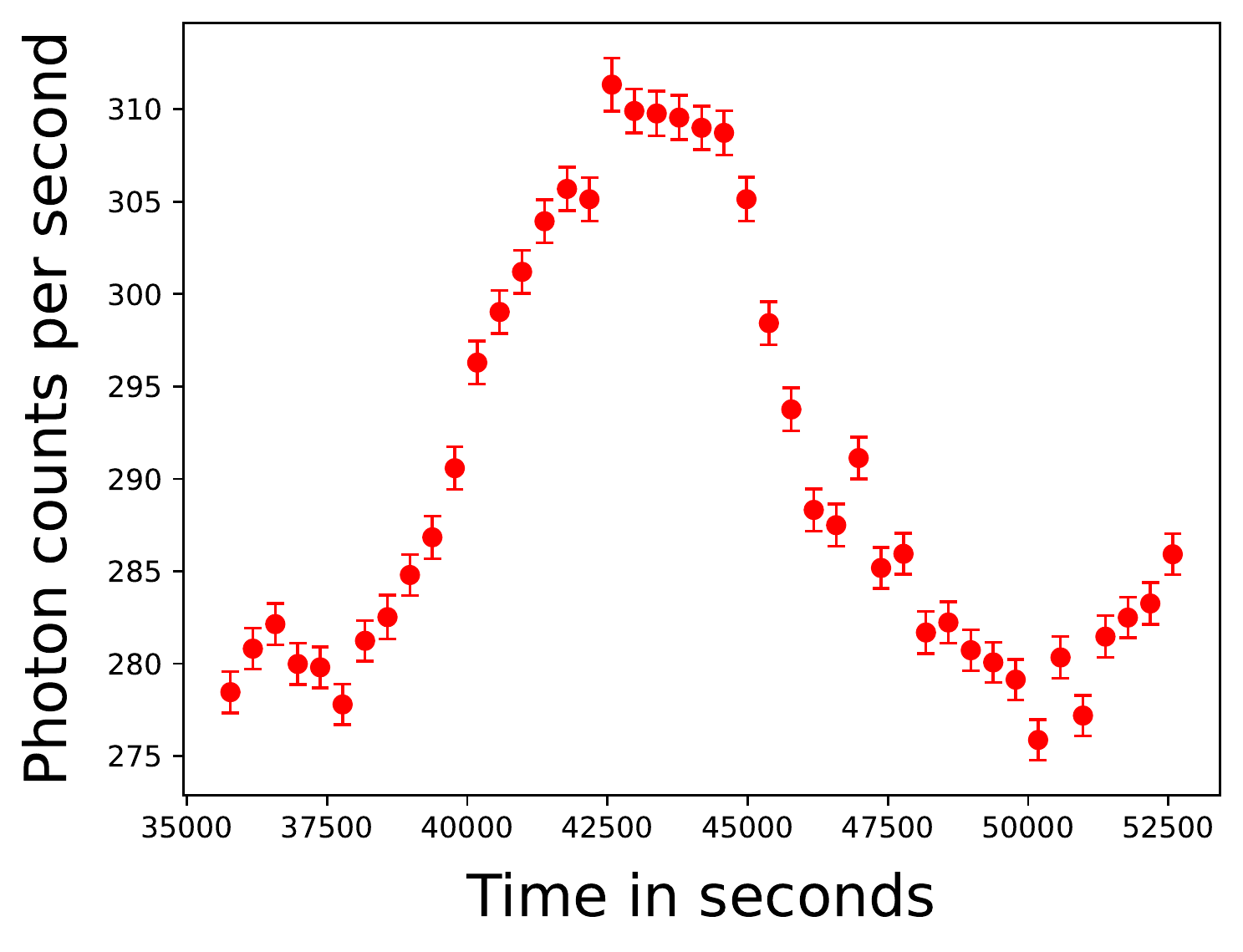} &
(c) \includegraphics[width=48mm, angle=0]{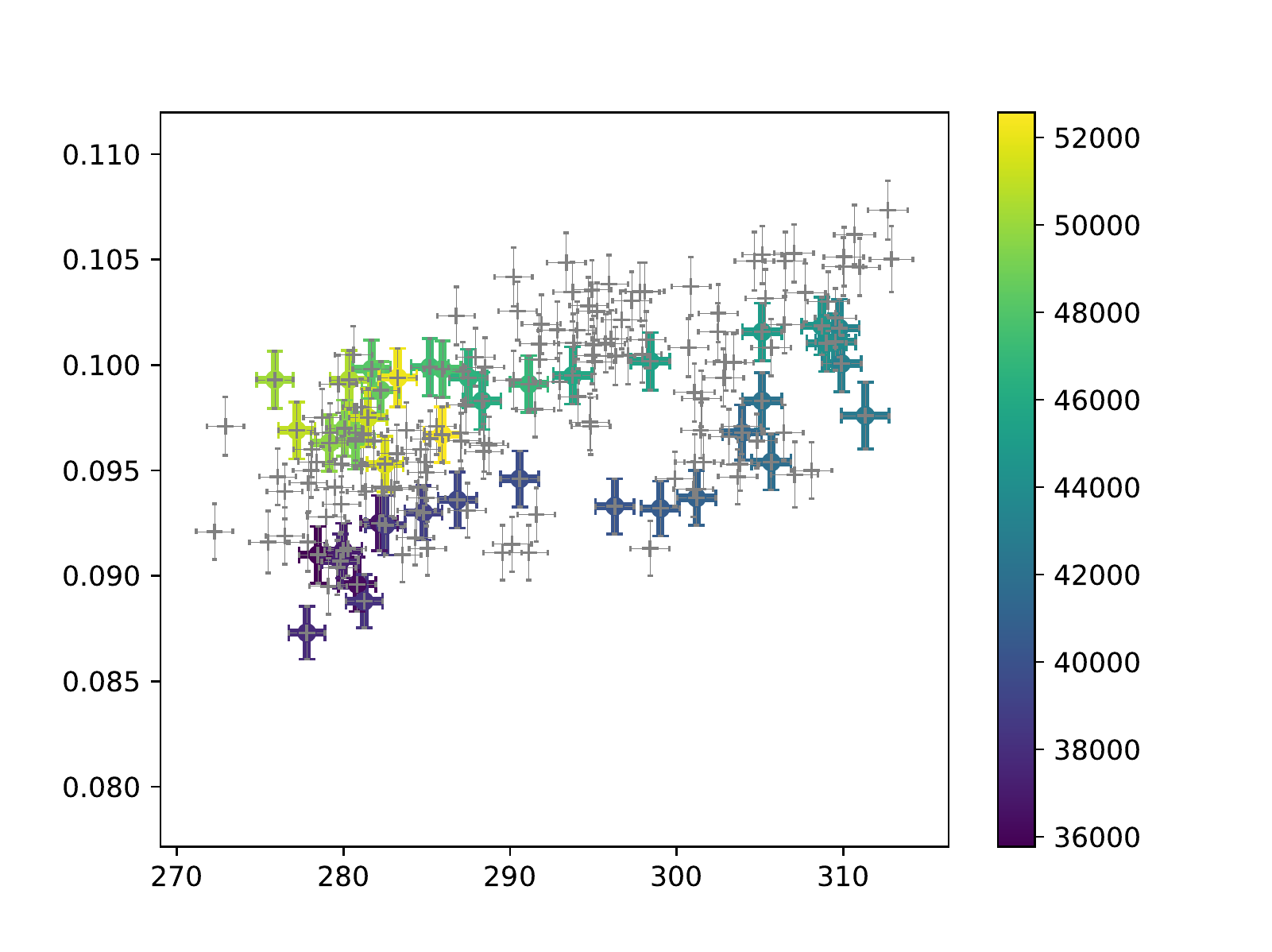} \\

\end{tabular}
\caption{  Obs. ID 0136541001 : (a) the light curve  with the considered flare marked by vertical lines; (b) the light curve of the flaring part; (c) HR--I diagram of the flaring portion of the light curve plotted in colour over the whole light curve HR--I diagram (Figure \ref{fig 1}, online element 5, panel (c)) {with colorbar showing increasing time in seconds}, plotted  in gray.}
\label{fig 5}
\end{figure*}

 The flux is highly variable: from 248 to 284, from 22 to 30 and from 272 to 313 cts/s in the energy ranges of 0.3 - 2.0 keV, 2.0 - 10.0 keV and 0.3 - 10.0 keV, respectively.
The time lag of hard photons derived from the DCF  is -1.11 ks, but the gaussian fitting looks poor with the peak of the gaussian not being consistent with the time corresponding to the highest DCF value. The inspection of the normalized S and H light curves do not provide clear clues where along the light curve the DCF delay originates.  
The HR-I diagram shows only a moderate harder-when-brighter trend. Also the HR is forming an anti-clockwise loop with respect to time in the HR-I plot. 

\cite{2004A&A...424..841R} analyzed only the flaring part of this observation and they reported a clockwise loop in the hardness intensity diagram. This is an interesting case as we see an anti-clockwise loop formation in the HR-I diagram for the entire observation. We have plotted the HR-I diagram for both the cases, the flaring part and the whole light curve, at the panel (i). This reversal of loop is an interesting behaviour revealed in this observation.

\subsubsection*{Obs. ID 0158970101}







This 47.5~ks observation depicts flux ranging from 282 to 327, 18 to 26 and 302 to 399 cts/s in soft, hard and total energy bands. {The fractional variability in soft and total energy band is almost equal while it is higher by 2\% in hard energy band.}
From the DCF, there is a clear positive lag in the observation of 2.56 ks, suggesting that the emission of hard energy photons precedes the emission of soft energy photons. This is also very evident in the normalized light curve with visible shift of hard and soft energy bands.  
In the HR-I diagram one clearly see a clockwise loop, but apparently there is no harder-when-brighter trend in the data.

\subsubsection*{Obs. ID 0150498701}








This 49.3~ks observation with clearly variable flux ranges from 483 to 714, 52 to 92 and 536 to 808 cts/s in soft, hard and total energy bands. 
The amplitude of fractional variability of this observations is approximately around 8\% in both soft and total energy bands, while it is much higher {and almost double} with a value of 17.61\% in the hard energy band.

There is a lag in soft photons of 1.71ks calculated from the Gaussian fitting on the DCF, but not clearly visible in the normalized light curves.
The HR-I diagram shows  a clear harder-when-brighter behaviour and forms a loop in clockwise direction with time.

\subsubsection*{Obs. ID 0162960101}









This is a 50.7~ks observation  with the flux ranging from 305 to 343, 32 to 41 and 339 to 381 cts/s in soft, hard and total energy bands.
There are relatively low variability amplitudes, with small differences between energy band (2.79\% for soft, 4.62\% for hard and 2.91\% for total), but still the fractional variability in hard state is significantly larger than the ones in other bands. The derived minimum flux variability timescale is ~2.61 ks.

The time lag is 0.04 ks which is a very small positive lag. 
The HR-I diagram looks scattered, with no HR variation with flux or time.

\subsubsection*{Obs. ID 0158971201}









This 66.1~ks observation has the highest flux as can be seen from the light curves. The flux value in total energy band goes as high as 1073 cts/s and the lowest flux value for this observation is approximately 691 cts/s. The flux ranges from 86 to 224 cts/s in the hard energy band and 603 to 856 in the soft energy band.
{This  observation shows also the highest amount of flux variability, with an amplitude in the hard state 22.85\%, while in the soft and the total energy ranges it is around 10\%. }
From the Gaussian fitting to the DCF, the time lag is calculated to be -0.38 ks.   The HR-I diagram shows  a complex structure consisting of multiple vertically shifted branches, with a clear harder-when-brighter trend in the full observation as well as in each individual branch. The HR is also highly varied with time reaching highest values of 0.26 in all analysed observational series..

\subsubsection*{Obs. ID 0153951201}







This 10.0~ks observation, the shortest observation in our set, is characterized with the flux ranges from 564 to 652, 58 to 73 and 623 to 725 cts/s in soft, hard and total energy bands, respectively.
The fractional variability amplitude is between 4\% (soft and total band) and 6.5\% for the hard band.

The time lag as calculated from Gaussian fitting of the DCF is small with the positive value of 0.23 ks. From the plots in figure 10e and 10f,  this lag is very small and can be considered as zero lag. The HR-I diagram does not show any clear structure and only shows a slight  harder-when-brighter trend.

\subsubsection*{Obs. ID 0158971301}









This 60.0~ks observation has a high range of flux of 551 - 797, 68 - 120 and 623 - 912 photon counts per second in soft, hard and total energy bands, respectively. There is also a flare visible in the light curve between 30 ks and 50 ks of the observation with the hard component preceding the soft one.
{In all the three energy bands, this observation has a high fractional variability amplitude of above 10\%. 
}
The time lag  in DCF fit is 1.31 ks showing that the hard photons precede the soft ones. The HR increases with increasing flux and forms clockwise loops in time along the general trend.  

\subsubsection*{Obs. ID 0302180101}









This 41.9~ks observation with the flux ranges from 443 to 517, 53 to 76 and 498 to 594 cts/s in soft, hard and total energy bands, respectively, {shows modest fractional variability of ~ 4\% for soft and total energy bands, and almost double value in the high band.} This is the observation with the longest minimum timescale of flux variation of 10.59 ks in our studied set of observations.  
The DCF time lag between hard and soft state is 0.90 ks. The HR-I diagram forms a bit complex structure with respect to time, showing an  average harder-when-brighter trend, as can be in seen in the online Figure \ref{fig 1}, (online element 12, panels (c) and (d)).
It is interesting to note in plot (e) that minima in the light curve in soft and hard bands are similar within the given band, but much different between the bands.

\subsubsection*{Obs. ID 0411080301}









This 69.2~ks observation has the largest average flux count in our set of data with the flux values in soft, hard and total energy bands ranging from 643 to 753, 109 to 150 and 752 to 902 cts/s, respectively.
{The fractional variability amplitude in the hard energy band is almost double of the value for soft and total energy bands.}
There is a slight (non-significant) positive lag between the hard and soft energy bands of 0.31 ks. The HR-I diagram depicts a trend supporting the harder-when-brighter behaviour. The HR is highly variable with respect to time and does not form any visible loop structure. 

\subsubsection*{Obs. ID 0411080701}









This 18.9~ks observation has small variability amplitude.  The flux ranges from 276 to 289, 18 to 21 and 295 to 309 cts/s in the soft, hard and total energy bands, respectively. 
There is a negative time lag of -0.94ks which means that the soft energy photons precede the hard energy photons. This lag is clearly visible in the DCF plot (f), but not particularly evident when inspecting the normalized light curves in the panel (e). In the HR-I diagram (c) a loop like structure can be noted, but an average spectral hardening with increasing flux is quite weak.

\subsubsection*{Obs. ID 0510610101}









This 27.6~ks observation has flux ranges of 244 to 256, 14 to 16 and 259 to 273 cts/s in soft, hard and total energy bands, respectively.
Just like the previous observation, this one also has a low fractional variability amplitude with a value of less than 1.5\% in soft and total energy bands. The value of variability amplitude in hard energy band is 3.36\%. The derived minimum flux variability timescale is ~4.69 ks.
A negative lag  of 0.24 ks is fitted to the DCF. The hardness-intensity diagram looks relatively flat.

\subsubsection*{Obs. ID 0510610201}









This is 22.7~ks observation with the flux ranges from 250 to 267, 15 to 17 and 265 to 284 cts/s in soft, hard and total energy bands, respectively.
{The small fractional variability amplitude is less for the soft and total energy bands but 3.91\% for hard energy band.}
There is a positive lag of 0.23ks between hard and soft photons obtained from the gaussian fitting to the DCF. This slightly positive lag is clearly visible in the DCF plot. The HR-I plot is quite flat and does not show any particular structure in time.

\subsubsection*{Obs. ID 0502030101}









 This is 43.2~ks observation with the flux ranging from 594 to 749, 47 to 73 and 642 to 822 cts/s in soft, hard and total energy bands, respectively.  The light curve shows nearly monotonic decrease throughout the observation.
The flux variability amplitude of this observation in soft and total energy bands is around 6.5\%. In the hard energy band, $F_{var}$ value is 10.19\%.  {There observation has a relatively long minimum flux variability timescale of ~9.60 ks.}
Also this observation shows in the DCF a negative lag, with the emission of soft photons preceding the emission of hard photons by 0.17 ks. From the plot of the Gaussian fitting over the DCF, this lag appears to be quite small and {can be considered as no lag}.
The HR-I diagram shows a non-monotonic harder-when-brighter behaviour, without forming any loop structure. 

\subsubsection*{Obs. ID 0670920301}









This short 16.2~ks observation presents  flux variations from 605 to 655, 41 to 48 and 646 to 702 cts/s in soft, hard and total energy bands, respectively. The light curve shows the flux to be regularly decreasing in time.
The fractional variability amplitude in all the three energy bands is around{ 3\%. 
The derived minimum flux variability timescale is quite short. 
}
According to the DCF fit the soft energy photons precedes the hard energy photons by 0.81ks.
The HR variability with respect to both flux and time is flat, without any regular  trend and no loop structures is visible in the HR-I diagram. 

\subsubsection*{Obs. ID 0670920401}








This short 18.0~ks observation depicts a low, continuously decreasing  flux,  ranging from 104 to 152, 4 to 7 and 109 to 159 cts/s in soft, hard and total energy bands, respectively. 
The fractional variability amplitude is really high in all  energy bands with values slightly above 10\%, which maybe possibly boosted by the measurements errors. This is the observation in which the lowest value of minimum variability timescale is calculated with the value of ~1.03ks. 
From the gaussian fitting to the DCF there is a negative lag of 0.21 ks.
The HR-I diagram is clearly flat with very low values of HR. 

\subsubsection*{Obs. ID 0670920501}









The light curve of this short 18.0~ks observation show variations from 550 to 590, 45 to 52 and 597 to 642 cts/s in soft, hard and total energy bands, respectively.
{The fractional variability amplitude is below 4\% in all the three energy bands.}
There is visible a very small positive lag of 0.51ks in DCF between the hard and soft energy photons.
The HR-I diagram shows a slight increase in hardness ratio with increasing flux. 

\subsubsection*{Obs. ID 0658801301}








This 29.0~ks observation has a flux range of 225 - 270, 31 - 43 and 256 - 312 cts/s in soft, hard and total energy bands, respectively.  
{The fractional variability amplitudes are ~ 6\% in the soft and total energy bands, while in the hard energy band it is 9.24\%.}
From the DCF plot it can be said that the time lag is consistent with zero. The HR shows a weak harder-when-brighter trend. 

\subsubsection*{Obs. ID 0658801801}








In the regularly growing light curve of this 33.6~ks observation one can see flux ranging from 196 to 243, 11 to 19 and 208 to 262 cts/s in soft, hard and total energy bands, respectively. The existing harder-when-brighter behaviour is nicely illustrated by the normalized light curves in the panel (e), with the rise in the hard band flux which is steeper than the rise in the soft band. 
The fractional variability amplitude in hard energy band is pretty high with a value of 14.02\% compared to the value in soft and total energy bands in which the values are around 6.4\%. 
The fitted positive lag in the DCF is 0.57ks, but there may be no lag in reality for this nearly monotonic changing light curve . 

\subsubsection*{Obs. ID 0658802301}









This 29.4~ks observation has flux ranges of 279 to 312, 17 to 23 and 298 to 334 cts/s in soft, hard and total energy bands, respectively. 
{This observation has low fractional variability amplitudes of ~4\% in the three energy bands. The minimum variability timescale is 1.64 ks.}
There is a significant negative time lag in DCF with the soft energy photons preceding the hard energy photons by 1.46ks. However, such delay is not visually pinpointed in the plot of normalized light curves due to large scatter in the hard band. The HR-I diagram does not show any trend with the flux.  

\subsubsection*{Obs. ID 0791780101}









This 17.5~ks observation is characterized with very small flux and is classified with our selected condition as non-variable. In the data the flux points range from 69 to 72, 5 to 6 and 75 to 78 cts/s in soft, hard and total energy bands respectively. The normalized light curve of hard band is only more scattered than the soft band, without any common trend between. {The fractional variability amplitude is around 0.5\% in soft and total energy bands and 2.21\% in hard energy band.}
{It is not easily understood that the formally derived minimum flux variability timescale for this case is so small of value 2.01 ks.}
The HR evaluated for this observation also does not show variability in time, but small growth trend with flux is not excluded. The DCF has been calculated and the Gaussian fitting has been done to this like all other observations to show negligible signs of correlated variability between the bands.  However, a small non-zero maximum near zero lag with $A \approx 0.2$ suggest small contribution of flux variability to the data scatter due to measurement errors.

\subsubsection*{Obs. ID 0791780601}









This very short 12.5~ks observation is one of two classified by the selected condition as non-variable. The measured flux points range for this observation in soft, hard and total energy bands is 357 to 371, 55 to 59 and 414 to 430 cts/s respectively.
The fractional variability amplitude in the hard energy band is 1.08\% while in soft and total energy bands it is below 0.8\%.  
From the DCF Gaussian fitting one finds a maximum with $A \approx 0.4$ and the time lag of -0.45 ks. In the normalized light curves a correlation between the hard and soft energy bands is evident. Thus some small variability must be present in these data, contrary to our clasification.The HR plots with respect to flux and time does not show any regular trend. 

\begin{table*}[htbp]
\caption{X-ray variability parameters of Mrk 421. In the successive columns we provide: Number (=the respective figure number); Observation ID;  $\bar{x}$ -- the average photons counts per second for a given observation; F$_{var}$ -- the fractional root mean square variability amplitude for soft, hard and total bands; $\tau_{var}$ -- the variability time scale, derived from the total flux LC;  \textit{A} -- the maximum value of DCF; $\mu$ -- a time lag at which DCF peaks; $\sigma$ -- a width of the fitted Gaussian function. All Observation IDs are arranged in a chronological order.}
\label{table 2}
\centering\small\setlength\tabcolsep{.225em}
\begin{tabular}{cccccccccc}\hline \hline
   & & & & F$_{var}$(\%) & & \\\cline{4-6}
Number & Obs ID     & $\bar{x}$  & Soft             & Hard             & Total            & $\tau_{var}$  & \textit{A}    & $\mu$   & $\sigma$\\
      &     & & (0.3-2.0 keV)    & (2.0-10.0 keV)        & (0.3-10.0 keV)      & (ks)         &                  & (ks)          & (ks)     \\ \hline
1 & 0099280101 & 317 & ~9.25 $\pm$ 0.07 & 16.49 $\pm$ 0.19 & 10.16 $\pm$ 0.07 & ~3.78 $\pm$ 0.41 & 0.99 & ~0.03   & 6.5 $\pm$ 0.0 \\
2 & 0099280201 & 130 & ~9.14 $\pm$ 0.08 & 10.44 $\pm$ 0.28 & ~9.17 $\pm$ 0.08 & ~4.81 $\pm$ 0.65 & 0.90 & ~2.44   & 9.8 $\pm$ 0.4 \\
3 & 0099280301 & 496 & ~3.03 $\pm$ 0.05 & 12.86 $\pm$ 0.13 & ~4.12 $\pm$ 0.04 & ~4.50 $\pm$ 0.54 & 0.77 & -0.19  & 2.3 $\pm$ 0.1 \\
4 & 0136540101 & 438 & ~5.09 $\pm$ 0.06 & 12.45 $\pm$ 0.19 & ~5.76 $\pm$ 0.06 & ~3.41 $\pm$ 0.33 & 0.88 & -0.21 & 1.6 $\pm$ 0.2 \\
5 & 0136541001 & 291 & ~3.41 $\pm$ 0.03 & ~6.61 $\pm$ 0.09 & ~3.62 $\pm$ 0.02 & ~4.19 $\pm$ 0.31 & 0.83 & -1.11   & 4.7 $\pm$ 0.7 \\
6 & 0158970101 & 329 & ~6.27 $\pm$ 0.07 & ~8.78 $\pm$ 0.28 & ~6.35 $\pm$ 0.07 & ~2.51 $\pm$ 0.24 &  0.67 & ~2.56   & 7.9 $\pm$ 0.7 \\
7 & 0150498701 & 694  & 7.61 $\pm$ 0.05 & ~17.61 $\pm$ 0.17  & ~8.29 $\pm$ 0.05 & ~2.96 $\pm$ 0.16 & 0.77 & ~1.71  & 7.4 $\pm$ 0.6 \\
8 & 0162960101 & 362 & ~2.79 $\pm$ 0.08 & ~4.62 $\pm$ 0.26 & ~2.91 $\pm$ 0.08 & ~2.61 $\pm$ 0.27 & 0.71 & 0.04   & 4.5 $\pm$ 1.1 \\
9 & 0158971201 & 850 & 9.51 $\pm$ 0.03 & ~22.85  $\pm$ 0.06 & 11.61 $\pm$ 0.02 & ~5.96 $\pm$ 0.52 & 0.94 & ~-0.38  & 2.2 $\pm$ 0.2 \\
10 & 0153951201 & 678 & ~4.30 $\pm$ 0.19 & ~6.41 $\pm$ 0.59  & ~4.50 $\pm$ 0.18 & ~7.02 $\pm$ 0.59 & 0.89 & ~0.23   & 3.1 $\pm$ 0.2 \\
11 & 0158971301 & 748 & 11.11 $\pm$ 0.04 & ~15.11 $\pm$ 0.11 & ~11.53 $\pm$ 0.04 & ~4.53 $\pm$ 0.33  & 0.96 & ~1.31  & 19.6 $\pm$ 2.2 \\
12 & 0302180101 & 537 & ~3.87 $\pm$ 0.02 & ~8.93 $\pm$ 0.07 & ~4.43 $\pm$ 0.02 & 10.59 $\pm$ 0.79 & 0.87 & ~-0.19   & 3.9 $\pm$ 1.2 \\
13 & 0411080301 & 822 & ~3.69 $\pm$ 0.05 & ~8.19 $\pm$ 0.14 & ~4.33 $\pm$ 0.05 & ~5.46 $\pm$ 0.53 & 0.88 & ~0.31  & 5.3 $\pm$ 0.3 \\
14 & 0411080701 & 302 & ~1.13 $\pm$ 0.05 & ~2.59 $\pm$ 0.22 & ~1.18 $\pm$ 0.06 & ~5.78 $\pm$ 0.45 & 0.62 & ~-0.94  & 5.9 $\pm$ 2.8 \\
15 & 0510610101 & 265 & ~1.21 $\pm$ 0.04 & ~3.36 $\pm$ 0.22 & ~1.32 $\pm$ 0.04 & ~4.69 $\pm$ 0.34 & 0.76 & ~-0.24  & 3.4 $\pm$ 0.5 \\
16 & 0510610201  & 276 & ~1.85 $\pm$ 0.05 & ~3.91 $\pm$ 0.21 & ~1.98 $\pm$ 0.04 & ~7.60 $\pm$ 0.69 & 0.84 & ~0.23  & 5.9 $\pm$ 0.5 \\
17 & 0502030101 & 703 & 6.38 $\pm$ 0.02 & ~10.19 $\pm$ 0.08 & ~6.67 $\pm$ 0.02 & ~9.60 $\pm$ 0.66 & 0.89 & ~-0.17  & 2.8 $\pm$ 0.6 \\
18 & 0670920301 & 673 & ~2.44 $\pm$ 0.09 & ~3.84 $\pm$ 0.36 & ~2.49 $\pm$ 0.09 & ~1.68 $\pm$ 0.15 & 0.77 &   -0.81 &  3.1 $\pm$  0.1 \\
19 & 0670920401  & 127 & 10.76 $\pm$ 0.11 & 11.69 $\pm$ 0.77 & 10.78 $\pm$ 0.12 & ~1.03 $\pm$ 0.07 & 0.76 & ~-0.21  & 1.6 $\pm$ 0.2 \\
20 & 0670920501 & 626 & ~1.64 $\pm$ 0.04 & ~3.78 $\pm$ 0.17 & ~1.76 $\pm$ 0.05 & ~4.91 $\pm$ 0.29 & 0.77 & ~0.15 & 1.3 $\pm$ 0.2 \\
21 & 0658801301 & 283 & ~5.97 $\pm$ 0.07 & ~9.24 $\pm$ 0.18 & ~6.38 $\pm$ 0.07 & ~1.92 $\pm$ 0.19 & 0.97 & ~0.05 & 10.6 $\pm$ 0.1 \\
22 & 0658801801 & 230 & ~6.13 $\pm$ 0.08 & ~14.02 $\pm$ 0.34 & ~6.65 $\pm$ 0.08 & ~2.74 $\pm$ 0.39 & 0.91 & ~0.57 & 4.3 $\pm$ 0.5 \\
23 & 0658802301 & 314 & ~3.16 $\pm$ 0.13 & ~3.92 $\pm$ 0.51 & ~3.21 $\pm$ 0.13 & ~1.64 $\pm$ 0.26 & 0.65 & -1.46  & 6.9 $\pm$ 0.4 \\
24 & 0791780101 & 76 & ~0.47 $\pm$ 0.12 & ~2.21 $\pm$ 0.43  & ~0.61 $\pm$ 0.11 & ~2.01 $\pm$ 0.23 & 0.18 & ~0.23  & 0.9 $\pm$ 0.2 \\
25 & 0791780601 & 420 & ~0.74 $\pm$ 0.09 & ~1.09 $\pm$ 0.24 & ~0.78 $\pm$ 0.08 & ~4.92 $\pm$ 0.56 & 0.46 & -0.45  & 1.1 $\pm$ 0.2 \\\hline
\end{tabular}
\end{table*}

\begin{figure*}[h!]
    \centering
    \begin{tabular}{lll}
(a) \includegraphics[width=50mm, angle=0]{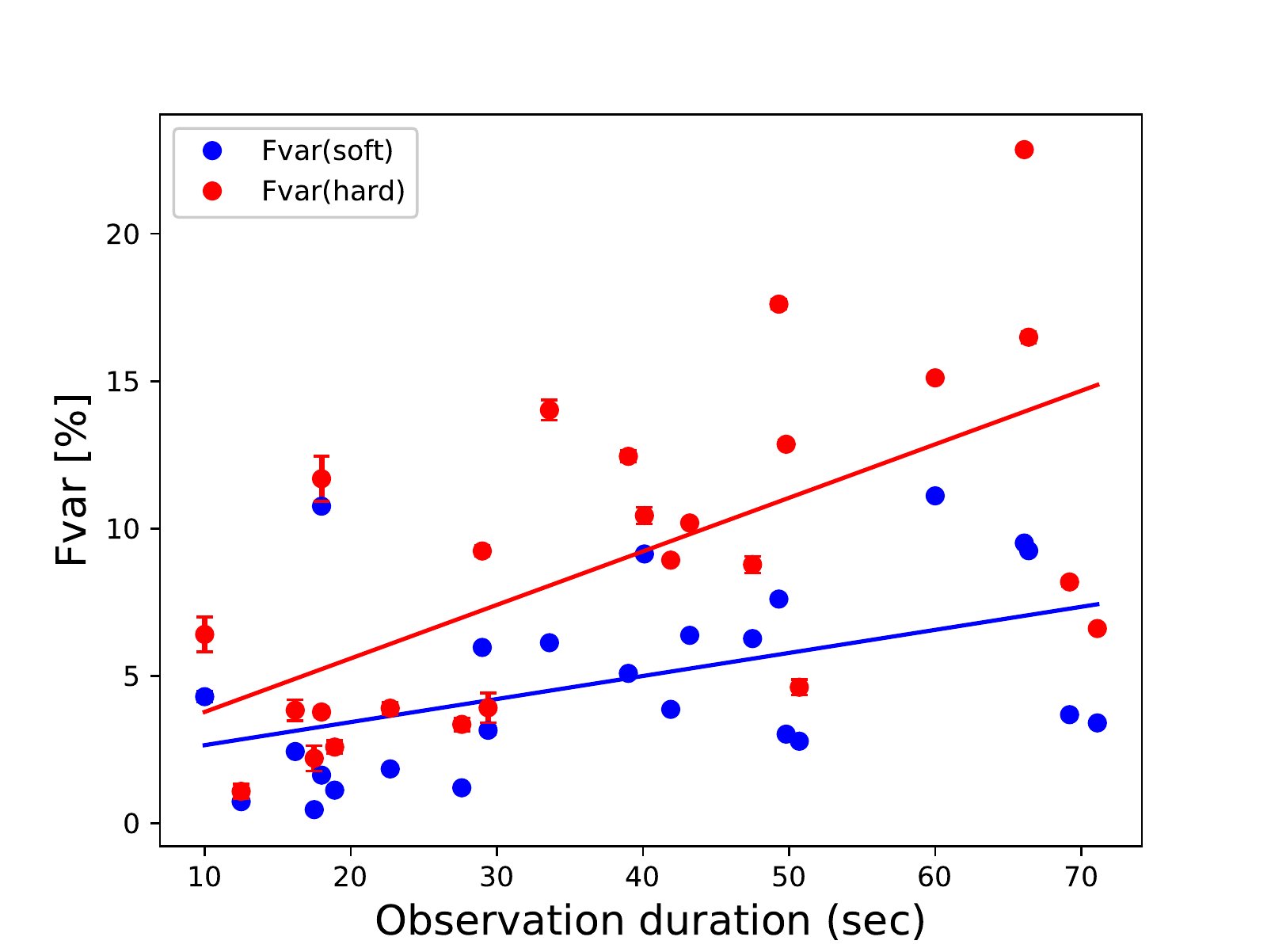} &

(b) \includegraphics[width=50mm, angle=0]{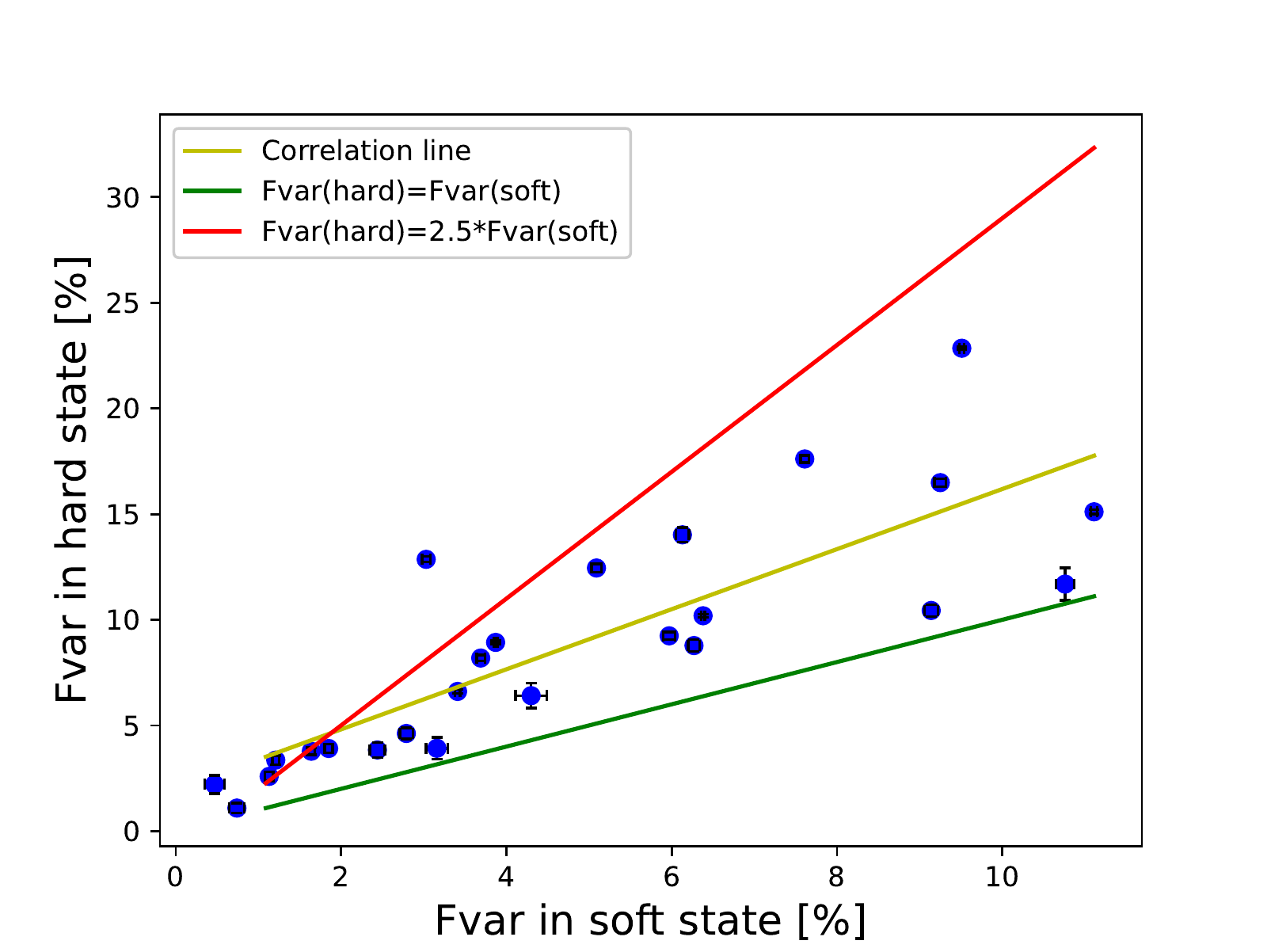} 
(c) \includegraphics[width=50mm,angle=0]{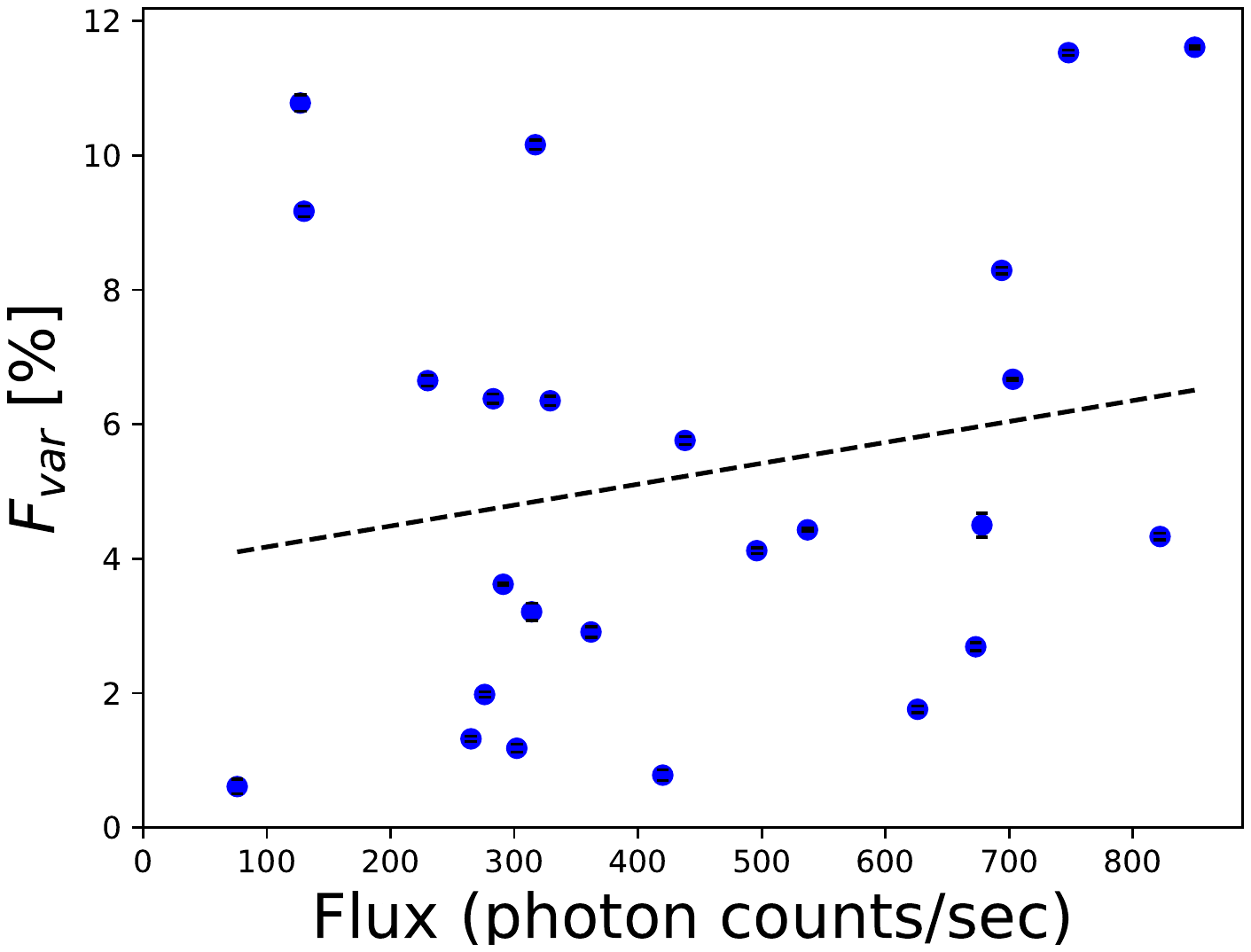}

\end{tabular}
\caption{(a) Fractional variability $F_{var}$ in the soft and  the hard energy bands plotted versus duration of observation, the fitted correlation lines are provided for both distributions. (b) $F_{var}$ in the hard versus the soft band. 
The lines $F_{var}(hard) = F_{var}(soft)$  and  $F_{var}(hard) = 2.5 F_{var}(soft)$ are provided for reference. (c) $F_{var}$ in total energy band versus flux with correlation line represented by dashed line}
\label{fig 26}
\end{figure*}

\begin{figure*}[h!]
    \centering
    \begin{tabular}{cc}     
    
(a)
\includegraphics[width=85mm, angle=0]{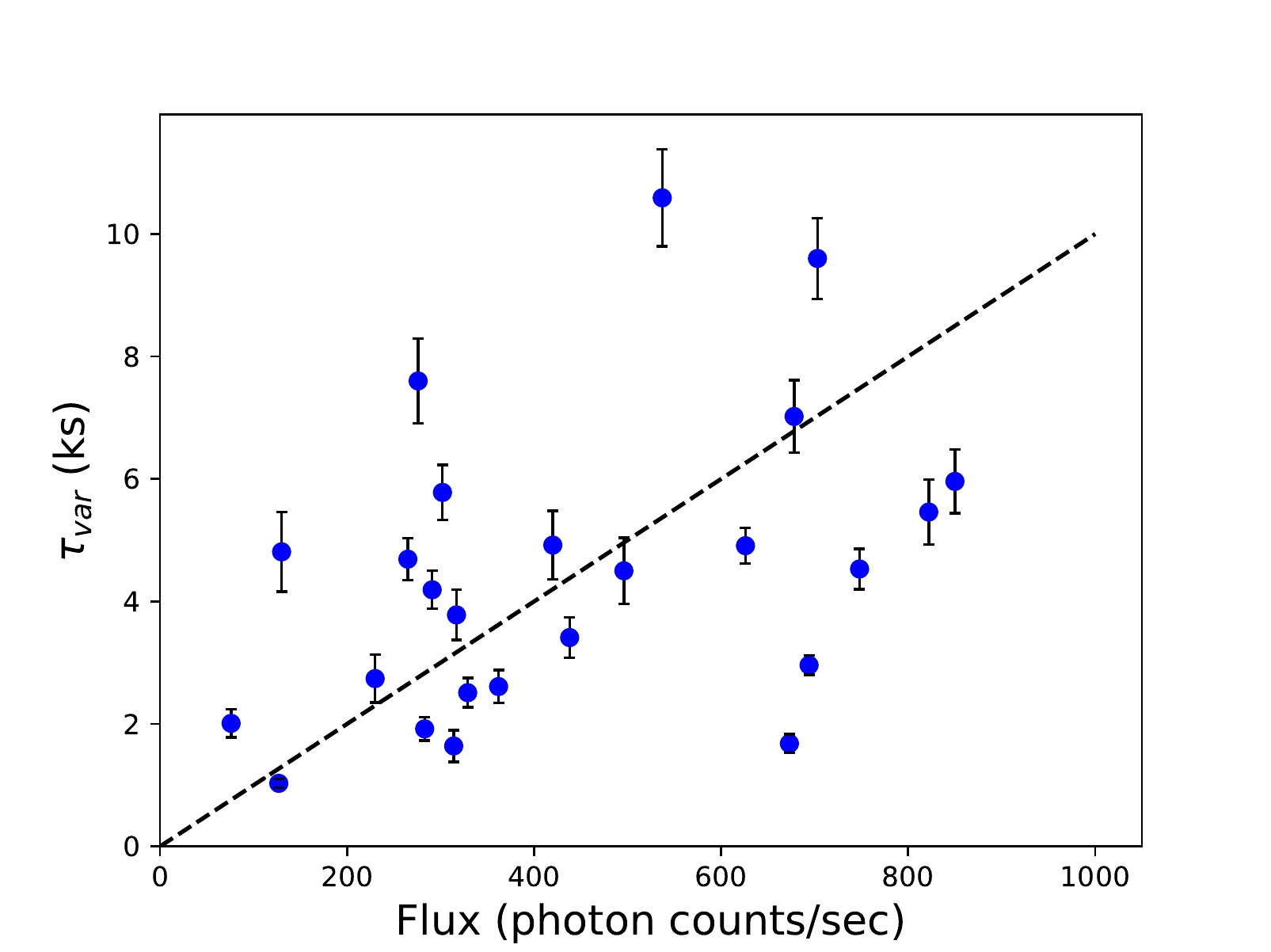} 

(b)
\includegraphics[width=77mm, angle=0]{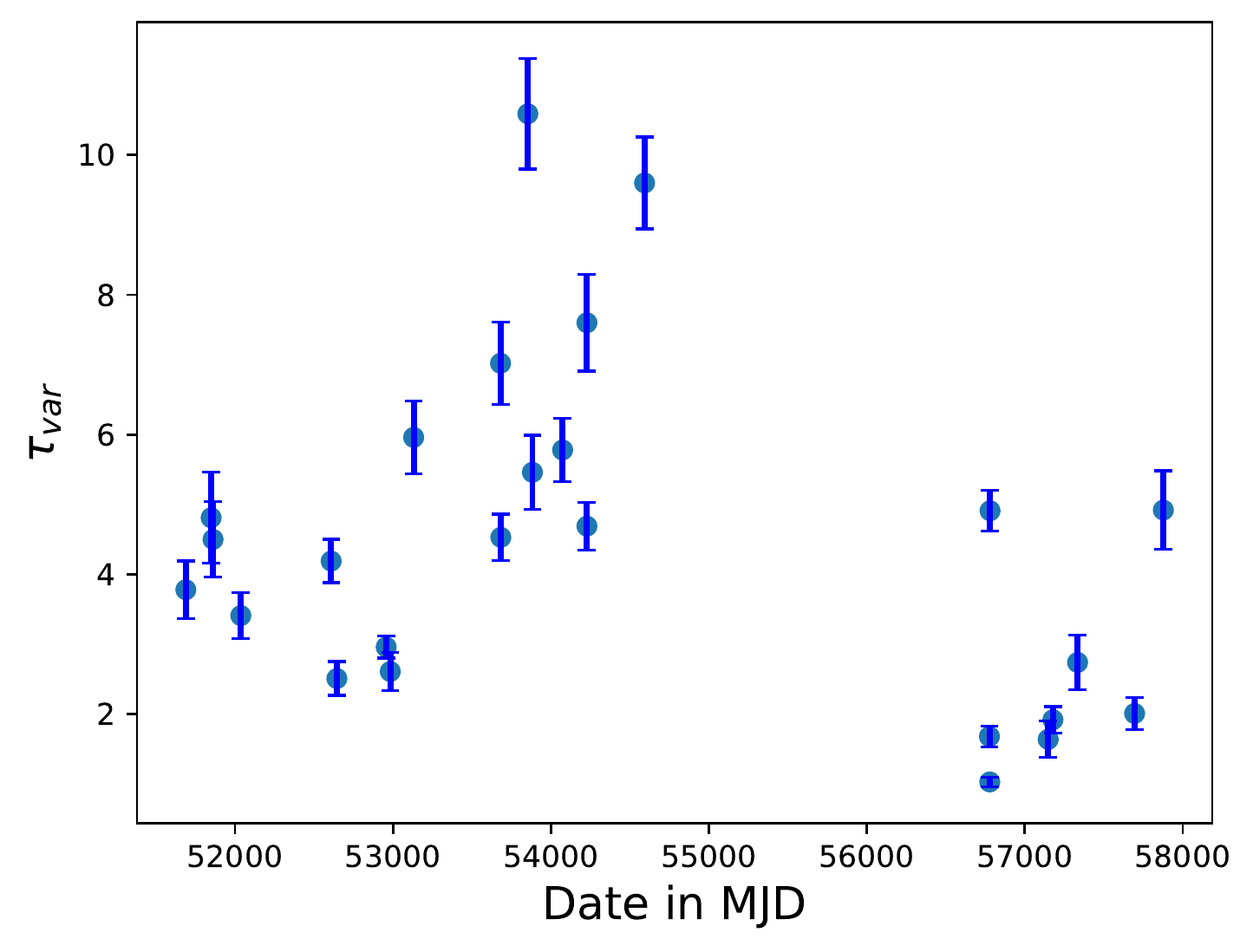} 

\end{tabular}
\caption{  (a)  A plot of $\tau_{var}$  vs. the mean photon flux $\overline{x}$  for all observations . An arbitrarily normalized dashed line $\tau_{var}$ = 4  ks {$\times$} Flux/(400 cts/s) is provided for a reference, showing the expected $\tau_{var}$ values for the same $\Delta x$ and $\Delta t$ used in all data sets. (b) The minimum variability time scale $\tau_{var}$  plotted against observation date given in MJD for all observations.}
\label{fig 27}
\end{figure*}

\subsection{Flux Variability}

For all 25 EPIC-pn {\it XMM-Newton} pointed X-ray observations, we study light curves in soft (0.3 -- 2.0 keV), hard (2.0 -- 10.0 keV), and total (0.3 -- 10.0 keV) energy bands, as presented in Figure \ref{fig 1} (the complete set of plots is provided in an online Figure set). Below we describe some observed general trends and findings from the analysis.

On visual inspection, we notice that all LCs  in the total energy band show substantial (23 observations) or weak (2 observations) flux variations on IDV timescales. To quantify these visual findings, we derived the fractional rms variability amplitude $F_{var}$ using equation (4) and the minimum two point variability timescale $\tau_{var}$ with equation (6),  for all the individual observation IDs. The results are provided in Table \ref{table 2}. {To make sure of the dependence of $F_{var}$ on the flux, the two parameters are plotted in figure \ref{fig 26}c for which the Spearman's rank correlation was evaluated to be 0.25. This shows a weak dependency of $F_{var}$ on flux. $F_{var}(total)$ and $F_{var}(soft)$ has a Spearman's rank correlation and pearson's correlation of approximately 0.5 with observation duration while $F_{var}(hard)$ and observation duration has a Spearman's rank correlation and pearson's correlation of approximately 0.6 (figure \ref{fig 26}a). These show the $F_{var}$ in different energy ranges have a moderate dependency on the duration of the observation.}

The often applied condition for the source variability, $F_{var}$ to be greater than $\rm{3} \times ({F}_{var})_{err}$, leaves only two observation IDs 0791780101 and 0791780601 as non-variable. All 23 other observational runs satisfy this condition for variability. Substituting these results into Eq. 12, we have estimated the source duty cycle DC in the total energy range (0.3 - 10.0) keV.  
Out of 25 pointed observations 23 showed variability on the IDV timescales yielding DC to be 96\%. {$F_{var}$ $>$ $\rm{3} \times ({F}_{var})_{err}$ is not the only criterion of judging variability in the light curves and failing this criterion doesn't necessarily mean that the source is not variable. It is also possible to fail to meet the criterion due to low photon count or shorter time duration.} Let us note that with visual inspection of the "non-variable cases", ID 0791780601 presented in Figure \ref{fig 1} (online element 25), shows some level of variability both from visual inspection of the LC and from the derived DCF shape with maximum correlation close to the zero delay between the hard and the soft LCs. Such correlation should not appear from fluctuations in sub-band of the source emission with a constant flux. 
For the second "non-variable" observation ID 0791780101 (Figure \ref{fig 1}, online element 24) hardly any clear signature of variability could be pinpointed in the light curve, however, inspection of the DCF plot shows again a "small maximum indicating weak hard/soft correlation close to zero time delay, what could be considered to suggest a weak variable signal added to LC fluctuations due to the measurement errors, in agreement with the red noise behaviour of Mrk 421. Therefore  the above DC value should be considered rather as the lower limit, with possible DC=98.6\%, or even 100\% if one believes our argument supporting a small amplitude variability in the LC of Observation ID 0791780101.  Due to these reasons we present all Obs. IDs with the plots of variability parameters. Our result here is consistent  with the Mrk 421 DC of 100\% and 84\% found using observations by {\it Suzaku} and {\it Chandra}, respectively \citep{2018MNRAS.480.4873A,2019ApJ...884..125Z}. 

The largest fractional variability amplitude in the total studied energy band is {11.6\% for the ID 0158971201} and the minimum $F_{var}$ = 0.61\%, is derived for the ID 0791780101. When analysing the variabilities in our selected hard and soft bands we note interesting trends presented in figure \ref{fig 26}. {$F_{var}(soft)$ and $F_{var}(hard)$ are highly correlated with each other with Spearman's rank correlation of approximately 0.9 and pearson's correlation of 0.82. For each plot in \ref{fig 26}, the correlation line is plotted using pearson's correlation values.}
When analysing the figure one should note, that particular points in  the presented distributions depend on our arbitrary split at 2 keV of the total band {into two energy bands}. This choice, characterized with having majority of photons in the soft band, leads to our HR values being much smaller than 1 and the variabilities of the total flux being very close to the soft band variability. 
the plot for a picture clarity.the plot for a picture clarity.the plot for a picture clarity.Thus the distribution of $F_{var}(total)$ is nearly identical to the one of $F_{var}(soft)$, with always $F_{var}(total)$ slightly greater than $F_{var}(soft)$), and we decided not to show it on the plot for a picture clarity. 
In figure \ref{fig 26} one may note that variabilities in the more energetic band are higher than the ones in the lower energy band, but also that on average in each energy band  $F_{var}$ grows with the observational time duration (figure \ref{fig 26}a), due to long wave light curve fluctuations  providing additional power in the long observation runs.  On the panel (b) of the figure we compare values of variabilities in both energy bands, showing that in all studied observations the variability in the hard band are substantially higher than the variability in soft band, but for the large majority of cases $F_{var}(hard) < 2.5 F_{var}(soft)$. The limits are: for the soft band, $F_{var}$ ranges from 0.47\% to 11.11\% for observations 0791780101 and 0158971301, respectively, while in the hard band it ranges from 1.09\% for observation 0791780601, to the value of 22.85\% for observation 0158971201. 

The extensively discussed here relation between hard and soft bands is not quite new in the literature. \cite{2018MNRAS.481.3563P} did a classification of AGNs on the basis of the X-ray variability and reported that for the radio loud AGNs variability is dominated by variability in hard band over variability in soft band. \cite{2001A&A...365L.162B} also pointed this out for one observation of Mrk 421 from XMM-Newton satellite.  Also the discussion of blazar power spectra in different energy bands by \cite{2020MNRAS.494.3432G} demonstrates domination of high energy variability at short IDV time scales for X-ray and gamma rays over optical and radio variabilities.

The minimum variability timescale in the analysed  observations ranges from 10.59~ks to 1.03~ks, the longest being for the observation 0302180101 and the smallest one for the observation 0670920401. To study the relation of $\tau_{var}$ with the activity state of the source which is measured here by the mean flux for a given Obs. ID, we have plotted on figure \ref{fig 27}a {$\tau_{var}$ against mean flux for each observation}. The distribution is highly scattered without any clear trend. To help in the evaluation we also provide a (arbitrary normalized) dashed curve showing the expected relation for the same  time bin used in all observations and by assuming the same characteristic maximum flux  change between two successive data points in all observations. At figure \ref{fig 27}b a quite interesting structure of $\tau_{var}$ changes over 17 years with the date of observation is observed. One should note the significant regularity of these changes in the plot, where in small time ranges a scatter in $\tau_{var}$ is quite limited, while it significantly evolves up and down over the observational period. In our opinion the observed non-random structure reflects regular physical changes of the blazar emission zone.

\subsection{Spectral Variability}

\subsubsection{Hardness ratio}

\begin{figure*}[h!]
    \centering
    \begin{tabular}{ccc}

(a)
\includegraphics[width=80mm, angle=0]{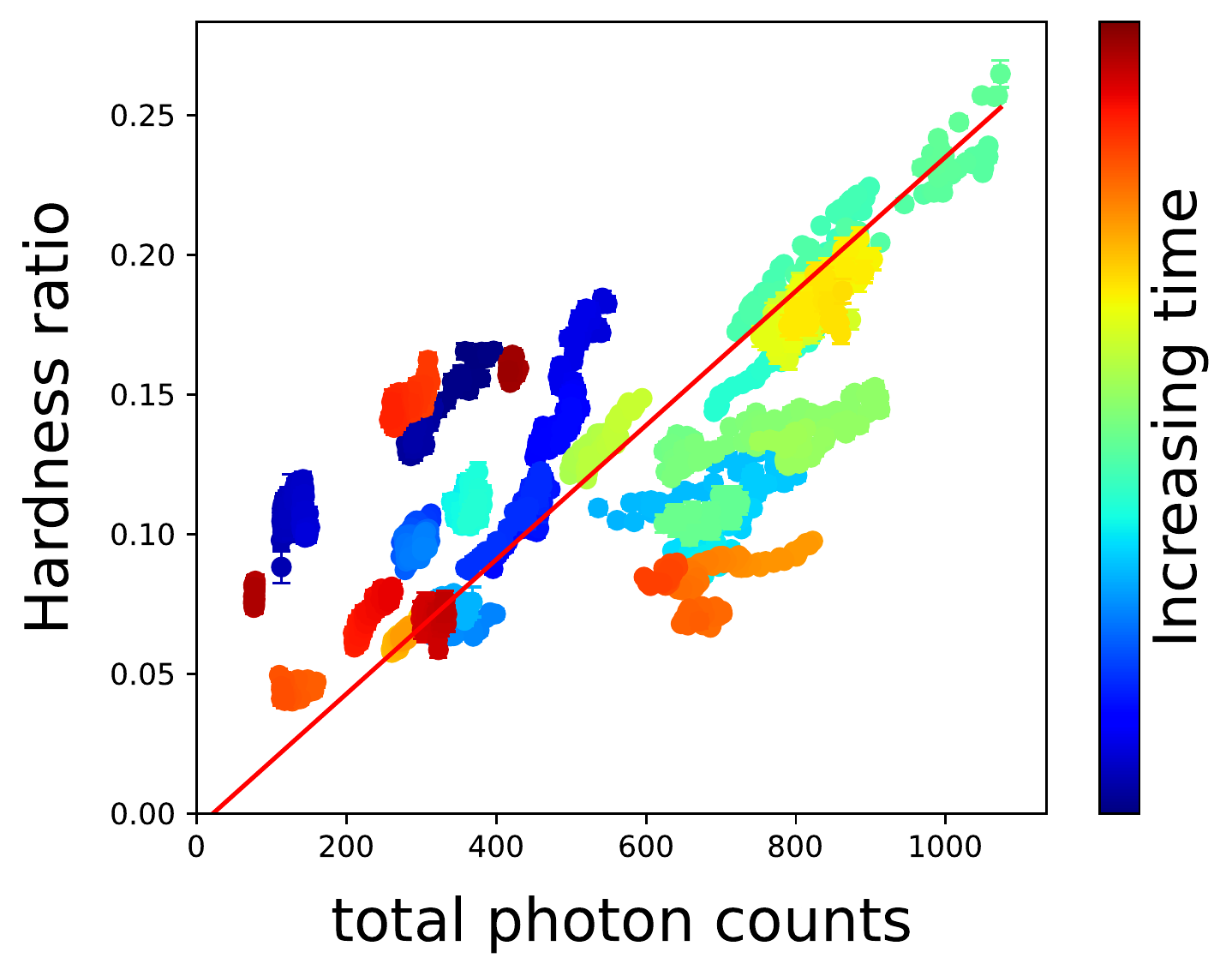} &

(b)
\includegraphics[width=80mm, angle=0]{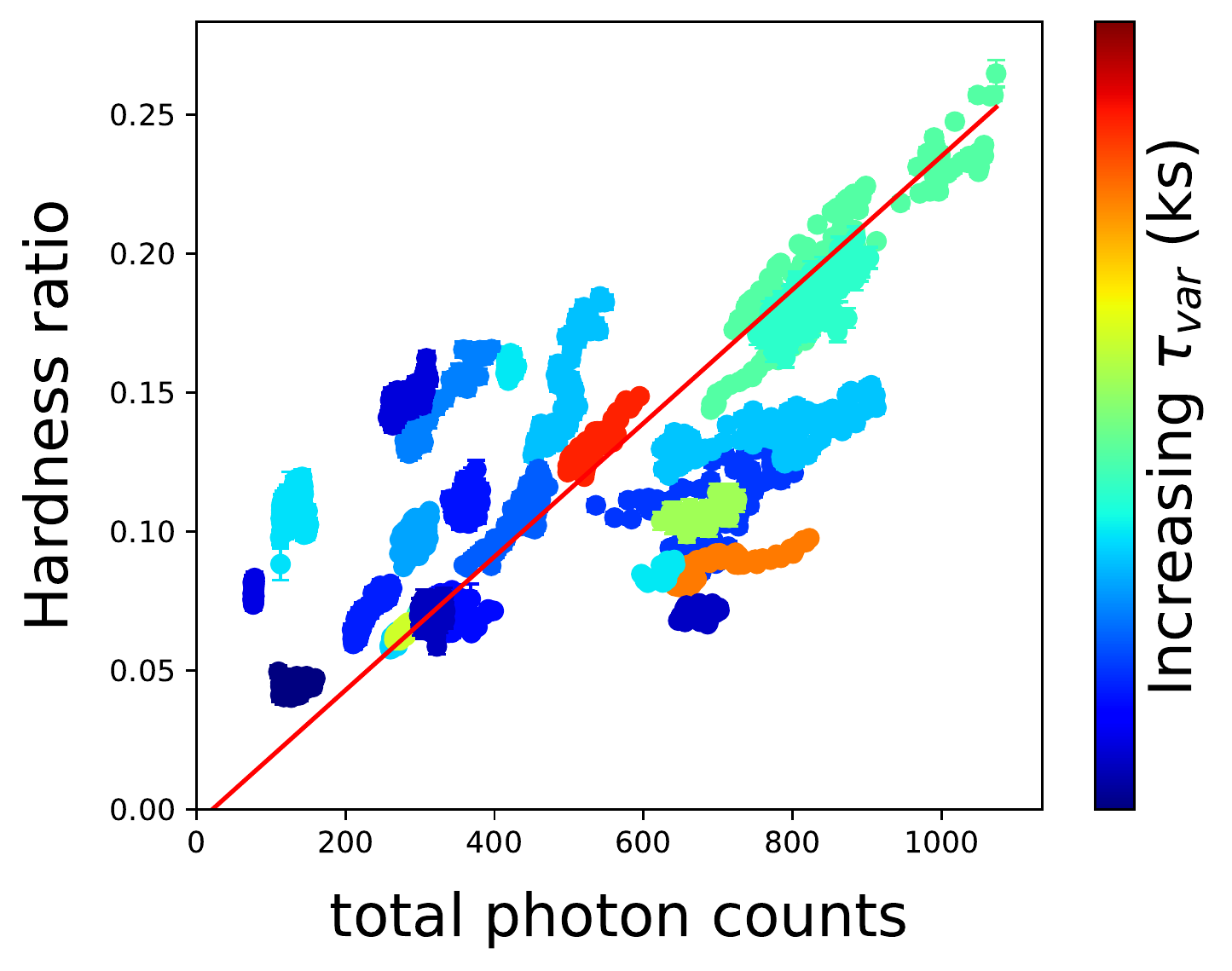} &

\end{tabular}
\caption{HR-I diagram of all plots combined into one with the color bar (a) depicting the {chronology of the observation starting with the first observation taken on 25 May 2000 and last observation taken on 4 May 2017}, (b) showing distribution of the derived $\tau_{var}$ within observations.}
\label{fig 28}
\end{figure*}

The harder-when-brighter behaviour is claimed to be a general property of HBL/HSP blazars \citep[e.g., see for review,][]{2020Galax...8...64G}. 
{The observations in this show more or less regular trends, with at} least half of the observations exhibiting the harder-when-brighter behaviour, but also including various zig-zags  with short vertical evolution paths or on average a constant value of HR. To compare these structures for all Obs. IDs we present the HR-I summary diagram for all observations in figure \ref{fig 28}, with a colour bar representing chronology of presented data. On the plot one can note a general harder-when-brighter trend,  {which we fit with a reference red line, but the structure seems to be more complicated. This red line is the line of best fit was obtained using the eyeball method.} If the line presents some reference trend for the source, then we note that at lower count rates, below ~500 cts/s, all data sets extend from or are situated above the  reference line to higher HRs. On the other hand for the higher count rates the HR-I distributions extend along the reference line or extend from the line toward lower HRs.

It is important to note that in many Obs. Ids  loop like structures are observed at HR-I plots, with both clockwise and anti-clockwise evolution in time. Occasionally both these orientations occur in a single observation.
The observation ID 0136541001 (figure \ref{fig 5}), taken on 1st of December, 2000, forms an overall clockwise loop in HR-I diagram. But on choosing only the flaring part of the light curve, the hardness intensity diagram depicts an anti-clockwise loop. This is in agreement with  \cite{2004A&A...424..841R} who showed an anti-clockwise loop for the flaring part of the hardness intensity diagram. This interesting behaviour depicts complexity of the emission processes. We note that observations forming a loop structure accompanies a time lag in hard energy or soft energy photons as calculated from DCF fitting, as. further discussed in the next section. 

\subsubsection{Discrete correlation function}

\begin{figure*}[h!]
    \centering
    \begin{tabular}{ccc}

(a)
\includegraphics[width=50mm, angle=0]{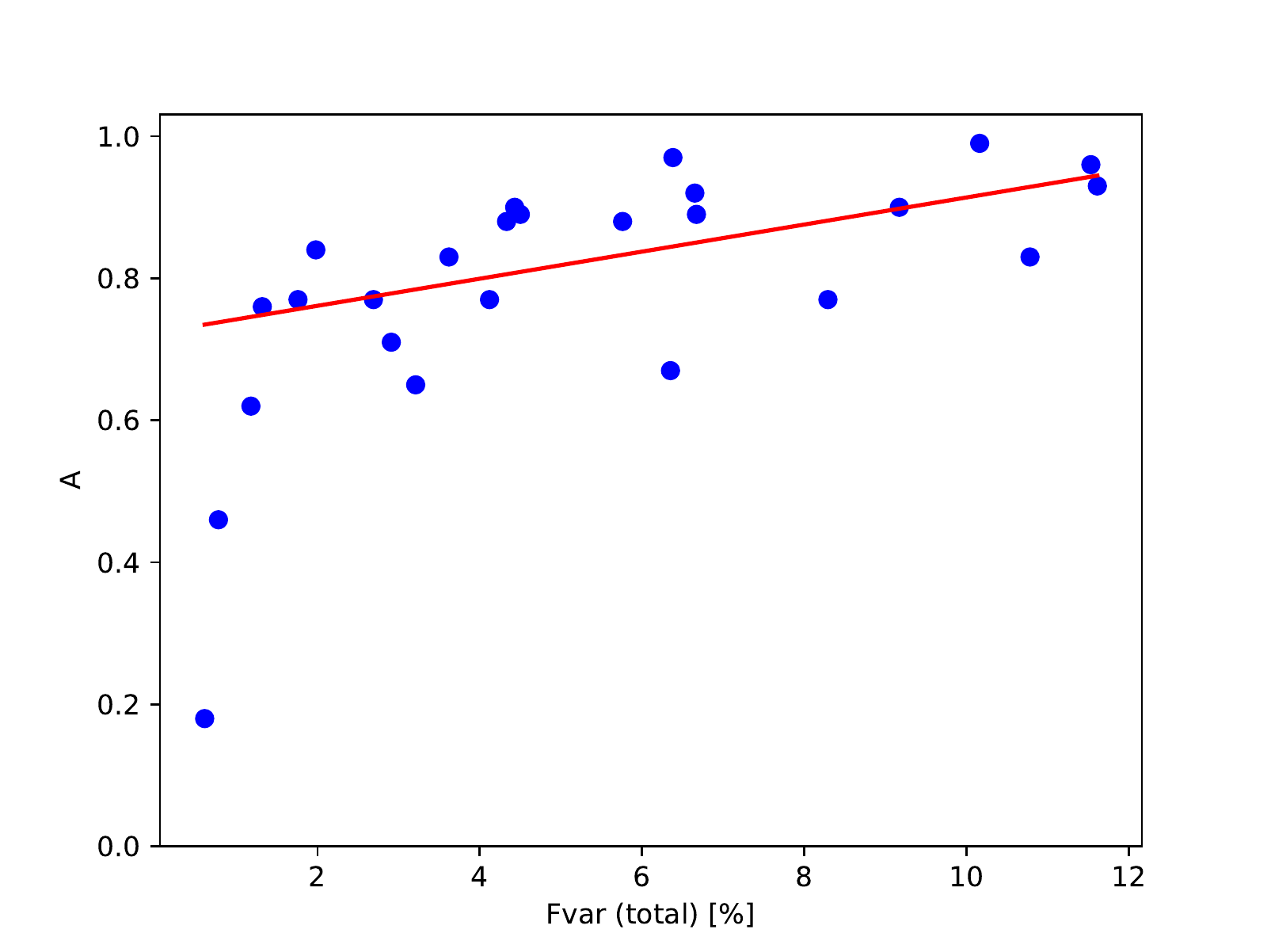} &

(b)
\includegraphics[width=50mm, angle=0]{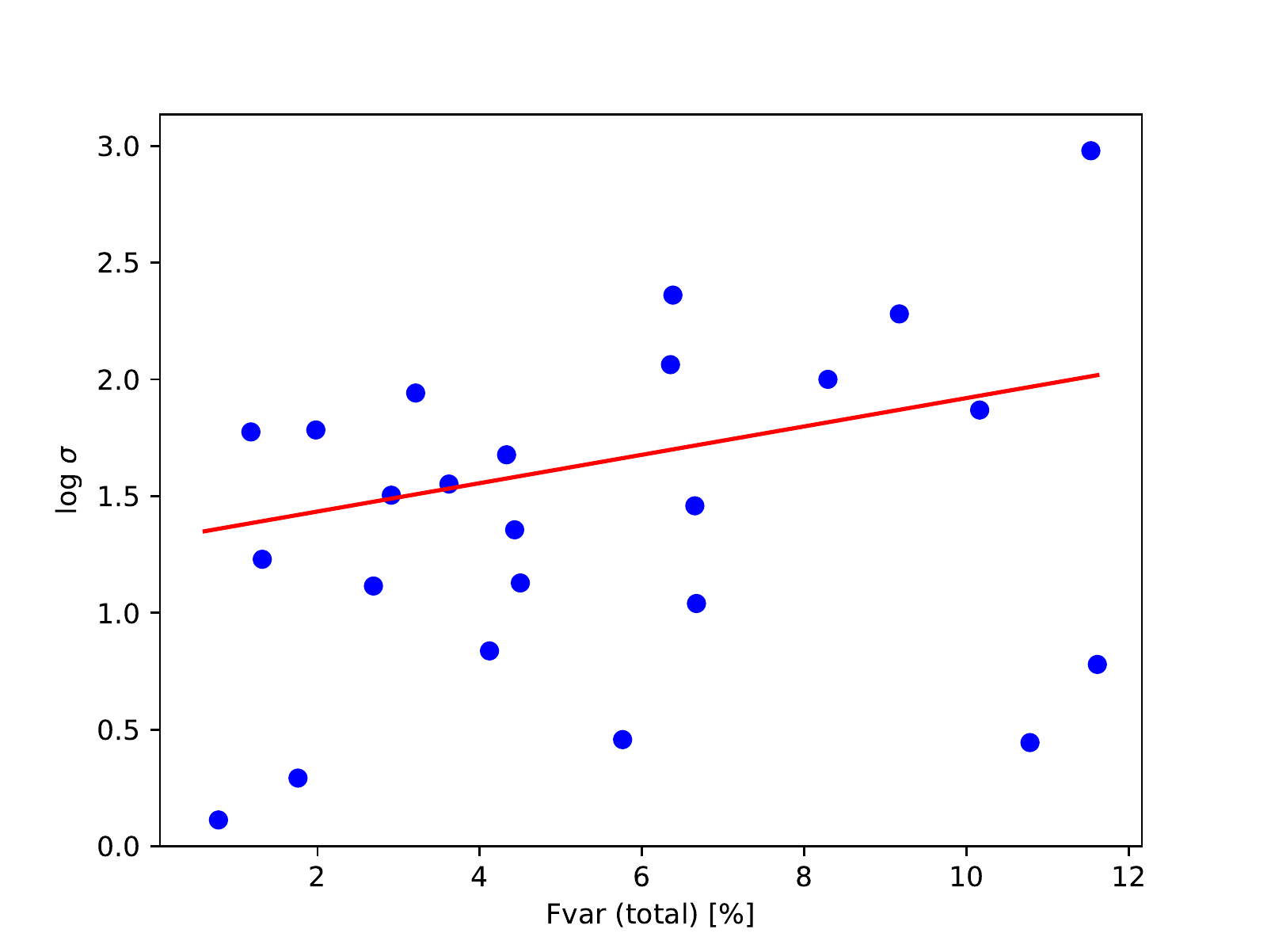} &

(c)
\includegraphics[width=50mm, angle=0]{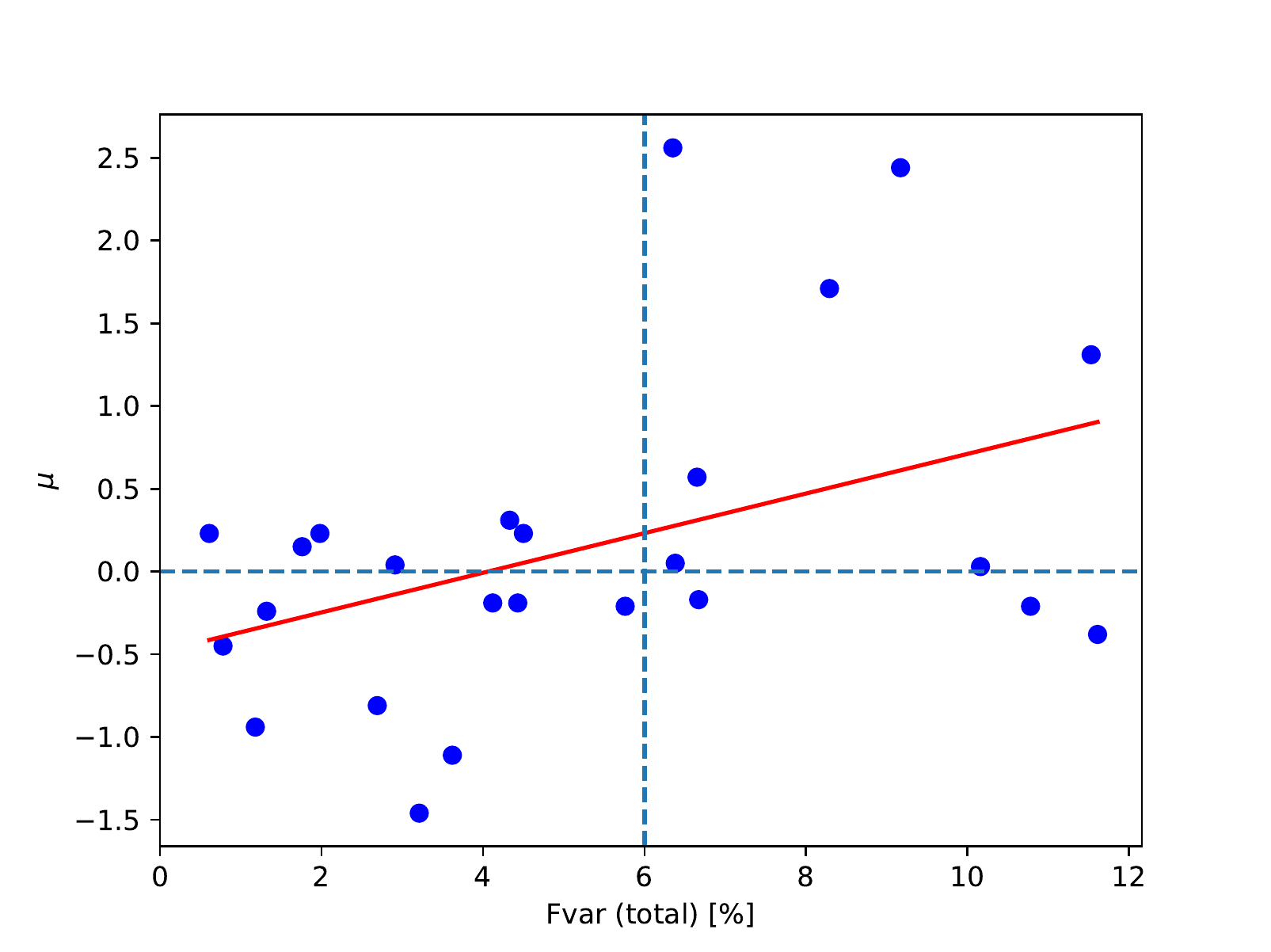} 

\end{tabular}
\caption{Fractional variability $F_{var}$ versus the parameters of the Gaussian fits of the DCF maximum: (a)  the amplitude A, (b) the width $\sigma$, (c) the time lag $\mu$. Red lines represent linear fits to the presented distributions, in without considering the two "non-variable" outliers with lowest $F_{var}$. The dashed lines in the panel (c) are for a reference, as explained in the text.}
\label{fig 29}
\end{figure*}

{All maxima in the DCF plots (Figure \ref{fig 1}, its online elements and figure \ref{fig 29}) have been fitted with the Gaussian function given by equation (11), with the fitted parameters presented in Table \ref{table 2}.} Then, these  parameters have been tested for correlations (or for non-random trends) with respect to the $F_{var}$ values in the total energy band (see figure \ref{fig 29}). Additional information resulted from the study of variations of the hardness ratio with respect to the flux intensity at the HR-I diagrams in figures 1-25 and compared all plots in figure \ref{fig 28}. Relation of another physical parameter characterizing the emission in each observation, also given in Table \ref{table 2}, was compared to the photon flux as shown in figure \ref{fig 27}a, and its evolution with time in figure \ref{fig 27}b.

\cite{2004A&A...424..841R} noted that DCF is not a robust way to predict lag in cases of highly variable light curves. While they have dealt with only the flaring parts in which the light curves were more or less symmetrical in time, we deal with the light curves of duration of more than 10 ks, which are often non-uniform and might consist of many flares or longer rise or decline trends. Thus,  to understand features in the light curves responsible for observed DCF lags in Figure \ref{fig 1} and the online Figure Set, we directly visually compare the normalized soft and hard  light curves, to reveal patterns or shifts between the fluctuating fluxes to understand origin of eventually occurring  non-zero time delays, $\mu$, appearing in the DCF analysis. A superposition of various lags in individual flares may lead to zero lag as also claimed by \cite{2006A&A...457..133F}. A positive value of the lag means that emission of photons in soft energy band is delayed with respect to emission in the hard band and it is also called as a "soft lag". When the soft energy photons precede the emission of hard energy photons then $\mu$ is negative and the lag is called the "hard lag".

Let us present main conclusions from these studies of DCFs and from comparison of the respective {\it normalized} light curves: 

1.) There is no statistical preference for the hard lag or the soft lag in the observations under study, with the soft lags observed in 13 observations, while the hard lags in 11 observations. The lag parameter $\mu$ for the observation 0791780101 is not taken into consideration because of not enough data points available for DCF modelling.

2.) {In our 25 observations, we do not notice a preference for lag in a particular energy band in the light curves, but involved light curve irregularities could mask small trends if present.} Occasionally we note situation with a hard band variation preceding the soft band one, or the contrary,  and in most cases it is always limited to a fraction of the entire observation. There are observations in which the trend of rise and fall in light curves is similar in the two energy bands, as observed in 0099280101, 0099280301, 0136540101, 0136541001 , 0158971301, 0302180101, 0411080301, 0510610201, 0670920401, 0658801801, 06588012301. 
There are cases of individual flares  with regular shapes in both bands showing no lag between frequencies; as observed in Obs. IDs 0099280101, 0136540101, 0162960101, 0158971201, 0158971301, 0302180101, 0411080301, 0510610201 and 0658801301. 
Also there are cases in which the trend of the light curves in two energy ranges is similar in a part of the observation while there is delay in the rise or fall of the photon counts in the other parts of the light curve in either of the energy band; as in observation 0099280201, 0158970101, 0162960101, 0411080701, 0502030101, 0670920501. 

3.) The registered larger time shifts $\mu$ of DCF maxima may be related with different evolution on longer time scales of photon fluxes in the different energy bands, and  such observations are also accompanied with occurence of  loop-like structures on the HR-I plots. Such behaviour is observed in observations 0150498701, 0158970101, 0158971201, 0158971301, 0136541001 and an indication of loop forming in observations 0099280201 and 0658802301. All the observations exhibiting a clockwise loops in the HR-I diagram are characterize with soft lags and, contrary, there are hard lags in observations exhibiting  anti-clockwise loops, as earlier observed in HBLs  by \citep{Zhang_2002, Zhang_2006, Fossati_2000}. But not all the observations in our data sample showing positive or negative lags are accompanied with  recognizable  loops in the HR-I diagrams.  

4.) {In figure 29, the gaussian fitting parameters of DCF maxima are plotted to study its relation with  $F_{var}$. Amplitude A is moderately correlated with $F_{var}$ with Spearman's correlation coefficient of approximately 0.7.}
Linear fits to these distributions are added to illustrate a general trend in each panel. In the panel (a) a moderate trend of growing the DCF maximum amplitude $A$ from 0.6-0.8 to 0.7-0.95 is observed, when $F_{var}$ grows from 1\% up to the maximum ~12\% (note, that two outliers with smallest $F_{var}$ were excluded from the fit here and in other two panels in this figure). Similar, {but a very weak trend with Spearman's correlation coefficent of 0.3} is observed for the width $\sigma$ of the fit in the panel (b).

5.)  In figure \ref{fig 27}(c), one may note a clear systematic trend for large negative or positive DCF lags between the two energy bands with respect to $F_{var}$.  Inspection of this largely scattered plot reveals an interesting structure with the all extreme negative (down to $<-1.5$ ks) $\mu$ values occurring  only at low variabilities,  $F_{var} \le 4$\%, while above 6\%  all the $\mu$ values deviating significantly from zero to the positive side, reaching up to +2.5 ks, appear. { Fisher's exact test was applied to our null hypothesis that the two parameters DCF lag and $F_{var}$ are independent of each other. The calculated Fisher's exact statistics value is ~ 0.2 and the null hypothesis is accepted. Thus, based on this test we cannot prove our speculation of dependency of positive or negative lag on the extent of variability. But on the other hand, there is a moderate correlation between DCF lag and $F_{var}$ with Spearman's correlation coefficient of 0.4 depicting a relation between large negative lags with low $F_{var}$ and high positive lag with high $F_{var}$.}
Thus a lag in soft photons more often is observed in more variable observations, while the lag in hard photons is dominant for low X-ray variabilities. 

\section{Conclusions}

In the present paper we re-analyzed a rich sample of 25 observations of TeV blazar Mrk 421 X-ray emission with the instrument EPIC-PN on board {\it XMM-Newton} attempting to do a multi-parameter detailed study of individual observations as well as to compare relations of the derived parameters for the full set of observations.  We used the obtained IDV light curves to study flux and spectral variations. 
A particularly interesting information was extracted from comparison of soft and hard X-ray sub-band variability properties by studying hardness ratio HR from the emission in the two sub-bands, the discrete cross correlation function DCF and comparing the normalized light curves in both energy bands. Verification of the accompanying HR-I diagrams and trends observed between them, as well as the trends in the derived minimum variability time scales for all data sets gives additional perspective in the study. Without simultaneous MWL data, besides a few speculations, we do not intend to model the studied emission processes, but rather we reveal significant constraints for any such modelling from a detailed single frequency band study to conclude:
 
\begin{enumerate}

\item[{1.}] The fractional variability displays clear evidence of large amplitude IDV in most of the observation IDs (23 out of 25  in all considered X-ray (sub-)bands. We found the IDV duty cycle to be 96\%, but some level of variability is noted in all data. 

\item[{2.}] The fractional variability amplitude depends on the studied X-ray energy range and it is always higher in the hard band than in the soft band (and in the total energy band).  In great majority of studied cases in the energy band selected by us, the fractional variability in the hard band is from $\approx 1$ up to 2.5 times higher that the variability in  the soft band.

\item[{3.}] The total energy weighted minimum variability timescales for all observation IDs occur in the range from 1.03 ks  to 10.59 ks. We note that this scale is not randomly distributed, but subject to regular trends in the 17 years of observations, possibly indicating underlying regular changes in the physical conditions  in the emission zone (like shifts of the emission zone along the jet with changing the magnetic field).

\item[{4.}]  The measured time lags between (0.3 - 2.0) keV (soft) and (2.0 - 10.0) keV (hard) bands from the DCF maximum fitting do not reveal any constant pattern. In majority of cases a very small time lag occur, while on inspecting the normalized light curves in hard and soft energy bands one can see positive, negative or unclear to pinpoint lags in different structures observed along the light curves.  
We also note that occurrence of long time trends between the studied energy bands result in time lags derived in the DCF analysis, even if short individual  flares in the high and soft sub-bands perfectly coincide in time. 

\item[{5.}] The HR analysis for our soft versus hard bands shows the similar pattern as the light curve for most of the observation IDs, with emission in the hard energy band being always more variable than in the soft band,  with the harder-when-brighter average trend of HSP blazars confirmed in majority of observations of Mrk 421. 

\item[{6.}] In numerous observations one may note formation of clockwise and anti-clockwise loops in HR-I  diagrams. 
We observe an interesting feature of change in the direction of loop in the HR-I diagram for the short time flare with respect to the loop appearing in entire observation of 0136541001.  
One may note that  \cite{Zhang_2006} attempted to explain a direction of such loops on the basis of energy dependent acceleration and cooling timescales of the emission particles, by somewhat arbitrarily varying these parameters for hard and soft photons. We prefer to think about possible physical source for such observations related to the complexes of relativistic magnetic field reconnection regions and relativistic turbulence within the jet volume modulated by the fluctuating jet density (or shocks).  
 
\item[{7.}] We observe that occurrence of the big lags in soft or hard photons is {moderately} related to the degree of flux variability. {There are large negative lags (soft photon lead) in less variable observations while highly variable observations have show large positive lags (hard photons lead).} We also see in our data set that the observations forming a clockwise loop in the HR-I diagram has a positive time lag while those forming an anti-clockwise loop has a negative lag.

\end{enumerate}

\noindent
In our opinion the observations with described trends or correlations clearly oppose application of shock acceleration mechanisms for explanation of blazar synchrotron emission in the X-ray range. In shocks high energy particles require longer times for acceleration and our observed events with high energy emission preceding the low energy one is in contradiction to such model. Also, in the flares generated by shocks some degree of regular asymmetry between the rising and declining parts of the flare is expected, which is not observed in the data. Of course the shock could still be a modulating factor through its compressive effects for some other acceleration and emission processes occurring in the medium, but even such influence should be reflected in some systematics in the observations, which we do not pinpoint here. So, even without referring to MWL observations, {we suggest that processes involving (relativistic magneto-hydrodynamics) turbulence are required to explain the considered X-ray data contrary to one zone. This is contrary to the earlier usage of one zone \citep{B_ttcher_2010} and two zone \citep{2005ApJ...630..130B} for the modelling of blazar spectra.}

Similar questions can be asked when postulating the magnetic field reconnection processes as the emission sources. In such situation one could imagine  the hard band preceding or being delayed with respect to the soft band emission, but still symmetry of some individual flares is difficult to explain if the reconnection is not modulated by an external symmetric process. In our view such models could fit the varying emission details in different observations if the emission is generated by modulated ensemble of numerous reconnection processes operating with time scales much shorter than the flares we observe. Fluctuating Lorentz factors within the jet would further complicate the situation.

Of course any model intending to explain the blazar emission should be verified  with simultaneous MWL observations. However, any such verification  should consider (as far as possible) IDV time scales where individual emission acts can give insight into the physics of acceleration processes. As a rule there are significant numbers of free parameters in the blazar emission models and one should be careful while making any real physical modelling, if not limited to playing with the free parameters only.

\begin{acknowledgements}
\textit{Acknowledgements}
This research is based on observations obtained with XMM-Newton, an ESA science mission with instruments and contributions directly funded by ESA member states and NASA.  The work of ACG is partially supported by Indo-Poland project no. DST/INT/POL/P19/2016 funded by the Department of Science and Technology (DST), Government of India. HG acknowledges financial support from the Department of Science and Technology (DST), Government of India, through INSPIRE faculty award IFA17-PH197 at ARIES, Nainital, India. GB acknowledges the financial support by the Narodowe Centrum Nauki (NCN) grant UMO-2017/26/D/ST9/01178. We would also like to sincerely thank Lukasz Stawarz for his valuable suggestions and discussions about the work.
\end{acknowledgements}

\bibliography{new.ms}{}
\bibliographystyle{aasjournal}
 
\end{document}